\documentclass[useAMS,usenatbib]{mn2e}

\usepackage{graphicx}
\usepackage{amsmath}
\usepackage{amssymb}
\usepackage{color}
\graphicspath{{figures/}}
\usepackage{fixltx2e}  

\renewcommand{\mp}{m_\mathrm{p}}
\newcommand{\mpccm}{\mp\,\mathrm{cm}^{-3}}
\newcommand{\iccm}{\mathrm{cm}^{-3}}

\newcommand{\Msun}{\mathrm{M}_\odot}
\newcommand{\Msunperyear}{\Msun \mathrm{yr}^{-1}}

\newcommand{\changed}{}
\title[Jet-induced star formation in gas-rich galaxies]{
Jet-induced star formation in gas-rich galaxies}
\author[V. Gaibler, S. Khochfar, M. Krause, J. Silk]{
    V. Gaibler$^{1,2}$\thanks{E-mail: v.gaibler@uni-heidelberg.de}, 
    S. Khochfar$^{2}$, 
    M. Krause$^{2,3}$ and
    J. Silk$^{4,5}$ \\
    $^{1}$Institut f\"ur Theoretische Astrophysik, Zentrum f\"ur Astronomie der Universit\"at Heidelberg, Albert-Ueberle-Str. 2, 69120 Heidelberg, Germany \\
    $^{2}$Max-Planck-Institut f\"ur extraterrestrische Physik, Gie\ss{}enbachstra\ss{}e,
    85748 Garching, Germany\\
    $^{3}$Excellence Cluster Universe, Technische Universit\"at M\"unchen, Boltzmannstra\ss{}e 2, 85748, Garching, Germany \\
    $^{4}$Institut d'Astrophysique, 98 bis Boulevard Arago, 75014 Paris, France \\
    $^{5}$Department of Physics and Astronomy, Homewood Campus, The Johns Hopkins University, Baltimore MD 21218, USA \\
}
\begin{document}

\date{Accepted ???. Received ???}

\pagerange{\pageref{firstpage}--\pageref{lastpage}} \pubyear{???}

\maketitle

\label{firstpage}

\begin{abstract}
  Feedback from active galactic nuclei (AGN) has become a major component in simulations of galaxy evolution, in particular for massive galaxies. AGN jets have been shown to provide a large amount of energy and are capable of quenching cooling flows. Their impact on the host galaxy, however, is still not understood. Subgrid models of AGN activity in a galaxy evolution context so far have been mostly focused on the quenching of star formation. To shed more light on the actual physics of the ``radio mode'' part of AGN activity, we have performed simulations of the interaction of a powerful AGN jet with the massive gaseous disc ($10^{11} \, \Msun$) of a high-redshift galaxy. We spatially resolve both the jet and the clumpy, multi-phase interstellar medium (ISM) and include an explicit star formation model in the simulation. Following the system over more than $10^7$ years, we find that the jet activity excavates the central region, but overall causes a significant change to the shape of the density probability distribution function and hence the star formation rate due to the formation of a blast wave with strong compression and cooling in the ISM. This results in a ring- or disc-shaped population of young stars. At later times, the increase in star formation rate also occurs in the disc regions further out since the jet cocoon pressurizes the ISM. The total mass of the additionally formed stars may be up to $10^{10} \, \Msun$ for one duty cycle. We discuss the details of this jet-induced star formation (positive feedback) and its potential consequences for galaxy evolution and observable signatures. 
\end{abstract}


\section{Introduction}
\label{sec:introduction}

Large scale cosmological simulations often include AGN feedback in massive galaxies in order to avoid both further growth of massive galaxies via star formation and the presence of a young stellar population at low redshift \citep[e.g.][]{Croton+2006,Bower+2006,Hopkins+2006,Schawinski+2006}. This approach is quite natural since AGN activity is frequently observed in these galaxies \citep{Best+2005}. So far, the models focus on negative feedback (suppression of star formation) to quench cooling flows, remove potentially star-forming gas by expelling it from the galaxy, or heating the interstellar gas to high temperatures in the halo to prevent new stars from forming. In contrast to radiation-based ``quasar mode'' feedback at high accretion rates, ``radio mode'' feedback by jets is powered mechanically by the kinetic energy of the jet beam \citep{Scheuer1974,BlandfordRees1974}. For the case of cooling flows in clusters of galaxies and giant ellipticals, there also is meanwhile considerable evidence from observations of X-ray cavities that the energies deposited by the jets suffice in many cases to suppress the formation of cooling flows in the hot circum-galactic and intra-cluster media \citep[e.g.][]{Birzan+2004,Dunn+2005,Rafferty+2006}. Furthermore, theoretical studies \citep[e.g.][]{Zanni+2005,Gaibler+2009,ONeillJones2010} show a high thermalization efficiency for this interaction. 

In this paper, we examine the impact of jet feedback on the star formation in massive, gas-rich galaxies at high redshift ($z \sim 2 - 3$). It is not obvious why jet feedback, which acts in a negative manner on the hot and dilute circum-galactic X-ray gas, should also act similarly on the interstellar medium (ISM), since this is a multi-phase medium with much higher densities and a pronounced clumpy density structure in dynamic equilibrium between gas at various densities and temperatures \citep{AvillezBreitschwerdt2005}. An alternative option is the model of positive feedback, where jet activity actually triggers or enhances star formation in a galaxy \citep{Silk2005}. This is, however, generally applied on the galaxy scales and without explicit hydrodynamical modelling in large-scale cosmological simulations. 

{\changed
Observational studies on this issue so far are inconclusive. Locally, one third of the radio galaxies show bright emission from warm molecular gas and very low to moderate star formation rates \citep{Ogle+2007,Ogle+2010,Nesvadba+2010}. The jets seem to heat the cold molecular gas and keep it warm during the lifetime of the jet although the cooling times are much shorter than the jet lifetime -- potentially due to the action of shocks, turbulent dissipation or cosmic rays \citep{Ferland+2008,Ferland+2009,Guillard+2009,Papadopoulos+2010}. The jets in those sources may actually induce formation of molecular gas but not cause the formation of a detectable number of stars \citep{Nesvadba+2011b}. Shock-heated, dense and turbulent gas has also been predicted in multi-phase turbulence simulations for jet cocoons \citep{KrauseAlexander2007}. Positive feedback, on the other hand, is directly supported by only a few observations locally (Minkowski's object in \citet{Croft+2006}, Centaurus A in \citet{Mould+2000,Morganti2010}), at intermediate (PKS2250-41 in \citet{Inskip+2008}) and high redshift (4C 41.17 in \citet{Dey+1997,Bicknell+2000}). But several studies point towards a link between jets and star formation, such as a high incidence of recent star formation activity in compact radio sources \citep{Dicken+2012}, the detection of large masses of cold molecular gas in high redshift radio galaxies \citep[e.g.][]{Emonts+2011}, alignment of excited CO with jets \citep{Klamer+2004}, alignment of continuum emission \citep{McCarthy+1987,Chambers+1987} and a correlation of emission line gas \citep*{McCarthy+1991} with the radio source. Furthermore, several authors find evidence for young stellar populations in radio galaxies \citep{Tadhunter+2002,Wills+2002,BaldiCapetti2008,Tadhunter+2011} and their central regions \citep{Aretxaga+2001}, possibly accounting for a substantial mass fraction \citep{Holt+2007}. It is important to keep in mind that the circumstances of jet feedback vary largely throughout these studies since gas masses and the velocity dispersions are much larger at high redshift (a setting that we are addressing in this paper) than in the local Universe, and variations amongst radio galaxies of a certain redshift regime are large, too.

We are still in an early stage of understanding jet feedback, despite very detailed observational studies in this field. The lack of physical modelling and understanding of jet feedback is currently one of the major shortcomings of ``radio mode'' feedback scenarios on galaxy scales. Only recently, high resolution hydrodynamical simulations including a clumpy interstellar medium have become feasible \citep{SutherlandBicknell2007,Gaibler+2011,WagnerBicknell2011}. Indeed, these studies have shown that the multi-phase structure of the ISM causes the morphology and jet--gas interaction to be qualitatively different from simulations without the cooler and denser phases. \citet{AntonuccioDeloguSilk2008} and \citet{Tortora+2009} have studied the interaction of a powerful jet in 2D slab geometry with cool clumps of gas dispersed within a hot medium, finding only a slight increase in the star formation (10 -- 20 per cent) followed by a much stronger decrease (more than 50 per cent) within a few million years due to heating and ablation of the cloud gas. However, the 2D approach results in a different pressure evolution of the cocoon and a larger effective surface of the clouds than in 3D. \citet{WagnerBicknell2011} studied the interaction of a relativistic jet with a clumpy gas distribution in a galaxy's centre at high resolution. They found a considerable impact of the filling factor of the dense gas on the interaction efficiency and that the dispersal of gas at large velocities may actually inhibit star formation. Yet, their simulated time-scales of $\la 1$ Myr allow only a weak impact of cooling and the simulations do not explicitly include a star formation model.
}

\defcitealias{Gaibler+2011}{Paper I}
In our previous paper, \citep*[][hereafter \citetalias{Gaibler+2011}]{Gaibler+2011}, we found that the action of jets in a clumpy medium results in a complex interaction with both individual clumps and the average density distribution of the gaseous distribution and results in stochastic asymmetries of the jets as one signature of this strong interaction. The resulting blast wave in a gas-rich galaxy compresses structures to high density where cooling is highly effective and impedes the cloud's pressure support and thereby its adiabatic re-expansion. In this paper we extend our earlier study to include star formation. We will here focus on the impact of the jet on the star forming gas in the galaxy and its observable signatures.
We find that a strong increase in star formation rate (SFR) is expected with a clear morphological signature in the stellar distribution. While the total number of stars in a given galaxy probably is not changed much by this, it would still be expected to be observable while it is ongoing and can have a significant impact on the galaxy's evolution.

The paper is organized as follows: in Sect.~\ref{sec:model} we summarize the setup of the simulation and state the differences to the setup of \citetalias{Gaibler+2011}. The results are given in Sect.~\ref{sec:results}, separating three different stages found in the simulation. The results are then discussed in Sect.~\ref{sec:discussion}.

\section{The Model}
\label{sec:model}

\begin{figure*}
   \centering
   \includegraphics[width=0.41\linewidth]{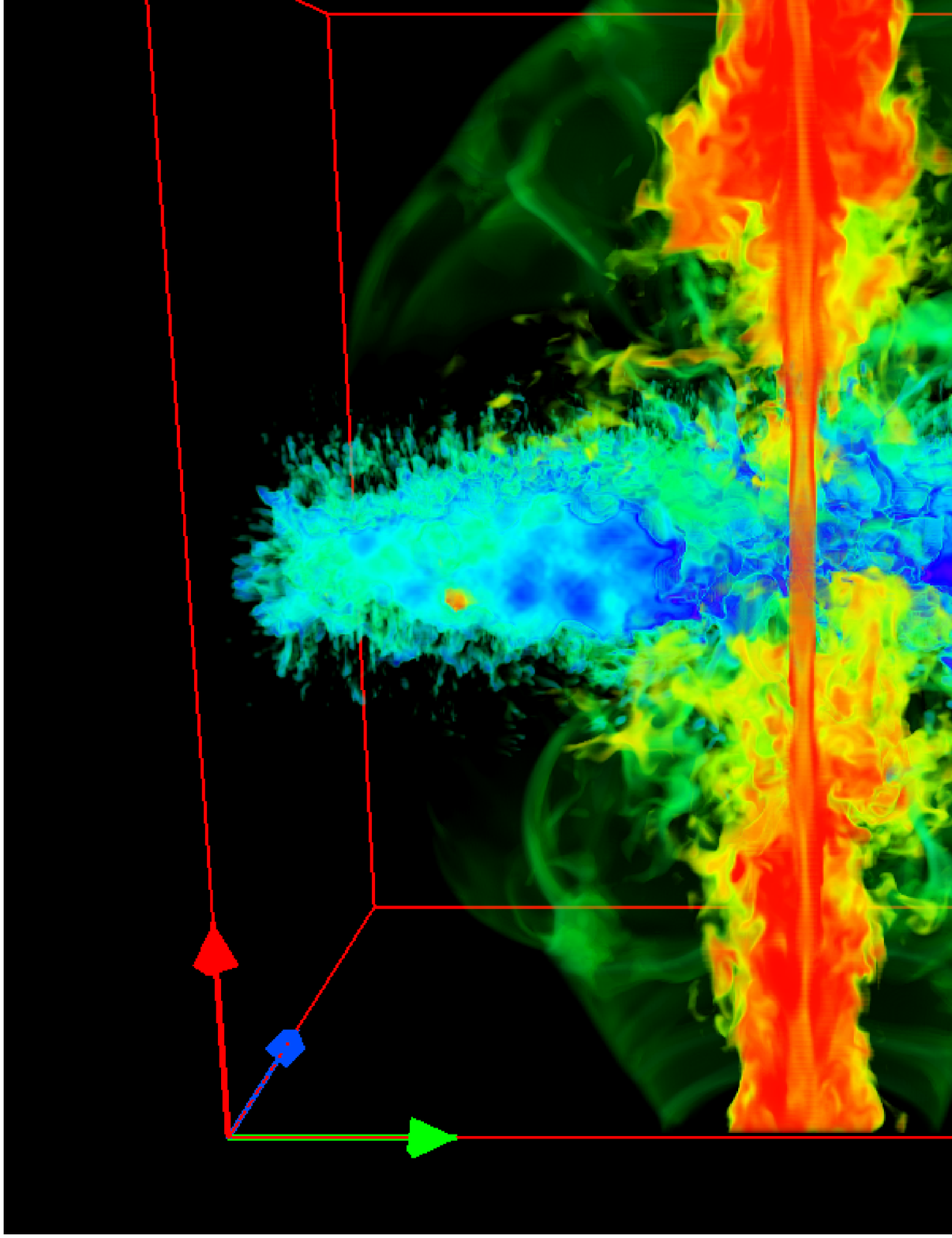}
   \includegraphics[width=0.41\linewidth]{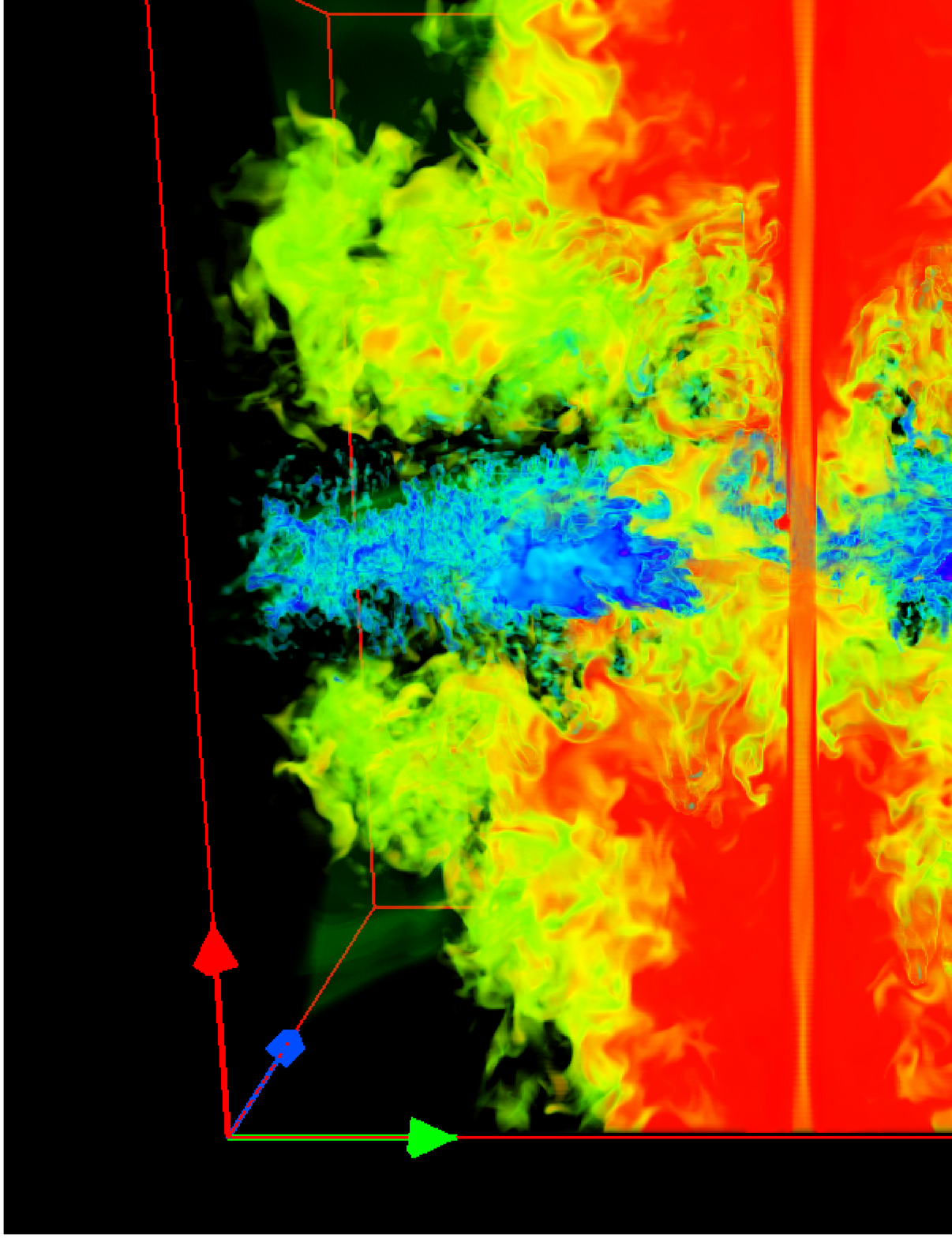}
   \caption{Density volume rendering of the central part of the domain (32 kpc box length) 
     at $t = 14$ (\emph{left}) and $t = 22$ Myr (\emph{right}). Only the $z > 0$ half is shown and densities close to the ambient X-ray gas are transparent to give a tomographic view. The colour bars show log $\rho$ in units of $\mpccm$.}
   \label{fig:density-view}
\end{figure*}
\begin{figure*}
   \centering
   \includegraphics[width=0.41\linewidth]{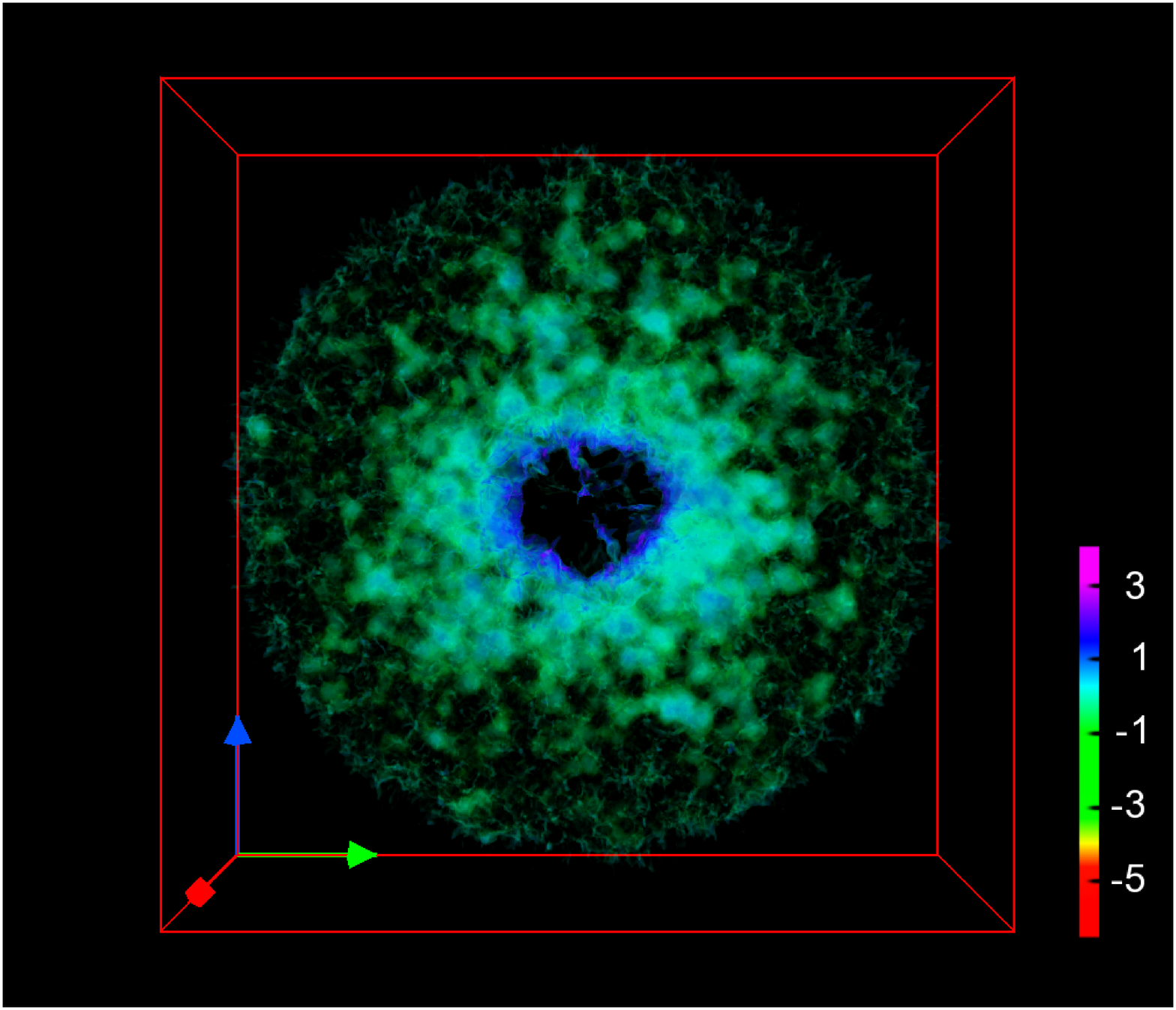}
   \includegraphics[width=0.41\linewidth]{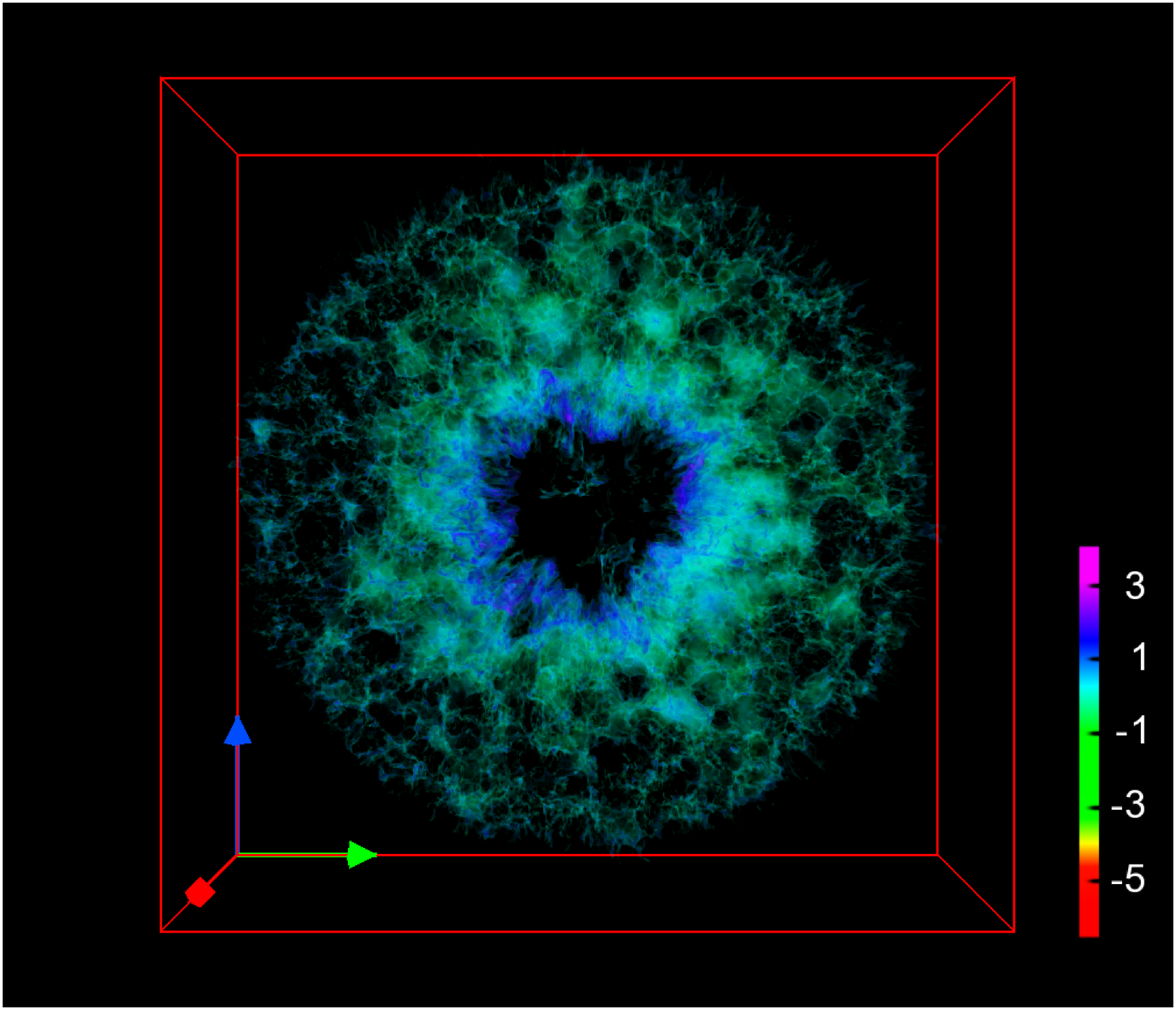}
   \caption{Density volume rendering of the high density gas in the disc (face-on view) at $t = 14$ (\emph{left}) and $t = 22$ Myr (\emph{right}). The box length is 32 kpc, lower density regions are transparent and the colour scale is the same as in Fig.~\ref{fig:density-view}.}
   \label{fig:only-dense-gas}
\end{figure*}
{\changed
We model the host galaxy as clumpy, disc-like distribution of gas with a log-normal probability distribution of the density (median $10 \, \mpccm$, mean $16.5 \, \mpccm$), an exponential radial profile with scale radius $R_0 = 5$ kpc and a sech$^2$ vertical profile with scale height $h_0 = 1.5$ kpc, reminiscent of a massive and thick galactic gaseous disc at high redshift \citep{Genzel+2006}. The width of the log-normal density probability distribution corresponds to a turbulence of Mach number 5 ($\sim 80$ km s$^{-1}$), which is similar to large evolving discs at high redshift \citep{FoersterSchreiber+2009}. The two-point structure of the density field is described by a Fourier spectrum that follows $E(k) \propto k^{-5/3}$ for large wave numbers $k$, but is damped towards smaller wave numbers to suppress inhomogeneities on scales larger than the disc scale height. The total mass of the disc is $1.5 \times 10^{11} \Msun$. The disc is embedded in a homogeneous hot atmosphere with a density of $0.05 \, \mpccm$ and a temperature of $1.15 \times 10^7$ K. The setup is almost identical to the one described in \citetalias{Gaibler+2011}; differences are noted in this section. The jet is implemented as cylindrical orifice of jet plasma (density $5 \times 10^{-5} \, \mpccm$, speed $0.8 \, c$) with the same pressure as the environment, a radius of $r_\mathrm{j} = 0.4$ kpc and an initial length in both directions of $3 r_\mathrm{j}$, respectively. 

We include radiative cooling by a tabulated cooling function for atomic processes down to $10^4$ K as \citet{SutherlandDopita1993} with a metallicity $Z = 0.5 \, Z_\odot$ \citep{Erb2008}, which is a very important effect in the disc due to the high densities and correspondingly short cooling times. No strong dependence on the metallicity or the exact form of the cooling function is expected since the mechanism found in our simulations can simply be understood by the fact that the cooling becomes more efficient (shorter cooling times) for denser gas. To avoid the need to include a fine-tuned feedback recipe to stabilize the cooling disc against the vertical gravity component we exclude gravity in the simulations and stabilize the disc by imposing a minimum temperature of $10^4$ K, which mimics a strong radiative heating from the stars that prevents any cooling below the threshold although this might still be expected in dense, shielded regions. 

Since cooling is included, no truly static setup of the disc is possible and we run control simulations of the disc without the jet for comparison in order to monitor the evolution of the undisturbed disc. In contrast to \citetalias{Gaibler+2011}, we use constant pressure for the initial conditions of the entire domain, however enforcing a pressure corresponding to a minimum temperature of $T_\mathrm{min}/\mu = 10^4$ K where necessary ($\mu$: mean particle mass in proton masses). Although the disc gas cools rapidly to $T_\mathrm{min}$ and drops out of pressure balance, it shows only little evolution in the fluid variables with little impact on the dynamics (however, resulting in a different jet asymmetry than in \citetalias{Gaibler+2011}). To allow for a more relaxed disc state, the jet of power $L_\mathrm{kin} = 5.5 \times 10^{45}$ erg s$^{-1}$ is only started at $t = 10$ Myr. The simulations are carried out using the RAMSES 3.0 code \citep{Teyssier2002}, a non-relativistic second-order Godunov-type shock-capturing adaptive mesh refinement code. The grid is refined to the maximum of $62.5$ pc cell size in all regions of interest. For the many dense clumps, the cooling length cannot be resolved and hence shells formed by shocks are expected to be thinner and denser in nature for these cases. We include tracer fields for both the jet plasma and the cool disc gas to follow its spatial distribution independent of mixing processes. Magnetic fields are not considered in this study. Our earlier magnetohydrodynamic simulations \citep{Gaibler+2009} showed that for the morphology of the jet cocoon at large scales they play an important role since they efficiently suppress Kelvin--Helmholtz instabilities at the contact discontinuity. Although the interaction of the jet with individual clumps of gas may be affected by magnetic fields and reduce the efficiency of ablation, this is beyond the scope of this paper and we leave it to a future study.  

In this study, we use a star formation model \citep{RaseraTeyssier2006}, where particles are generated in higher density regions of hydrogen number density $n_\mathrm{H} > n_\star$ ($\rho > 1.31 \, \mp \, n_\star$) according to $\dot{\rho}_\star = \epsilon \rho / t_\mathrm{ff}$, where $t_\mathrm{ff}$ is the local gas free-fall time and $\epsilon$ the star formation efficiency. This method is widely used in hydrodynamic simulations \citep[e.g.][]{Kravtsov2003,SpringelHernquist2003,SchayeDallaVecchia2008,DuboisTeyssier2008}, supported by observations \citep{KrumholzTan2007} and is able to recover the Kennicutt--Schmidt law \citep{Kennicutt1998} in simulations. After spawning a star particle, the corresponding gas is removed from the grid. We use two pairs of parameters $( n_\star = 0.1 \, \iccm, \epsilon = 0.025)$ and $(n_\star = 5 \, \iccm, \epsilon = 0.05)$ to check the impact of star formation being more strongly localized in high density regions. The efficiency parameters $\epsilon$ were chosen to get a SFR $\sim 150 \, \Msun$ yr$^{-1}$ similar to those observed in massive high-redshift galaxies \citep{Genzel+2010}. The higher star formation threshold can be regarded as being more realistic for our resolution, but shows more relaxation in the disc (which itself tends to decrease the SFR with time).
}

\section{Results}
\label{sec:results}

We here briefly summarize the formation of the blast wave as reported in \citetalias{Gaibler+2011}. 
The jets produce a blast wave in the central region of the disc since the density contrast between the jet and the dense gas ($\eta = \rho_\mathrm{j} / \rho_\mathrm{disc}$) is very strong ($\eta < 10^{-5}$), causing the jet head to stall and deposit its energy in mostly thermal form into the disc gas. This is also true at later times, when the jet breaks out of the disc, but since the density contrast is weaker, a stronger forward propagation results. Figs.~\ref{fig:density-view} and \ref{fig:only-dense-gas} give an overview of the evolution of the jet--disc interaction (see also fig.~3 of \citetalias{Gaibler+2011} and the supplementary movie there). Since the jet feedback is energy-driven, the large mechanical advantage \citep{WagnerBicknell2011} allows a much stronger effect then expected for a momentum-driven jet feedback model \citep{KrauseGaibler2010}. The early evolution hence is dominated by the effects of the blast wave rather than direct interaction of the jet beam with the ISM clouds, but the dense clumps inhibit the propagation of the jet and thus help to deposit the jet energy closer to the centre of the galaxy, thereby feeding the blast wave.

\begin{figure}
   \centering
   \includegraphics[width=0.9\linewidth]{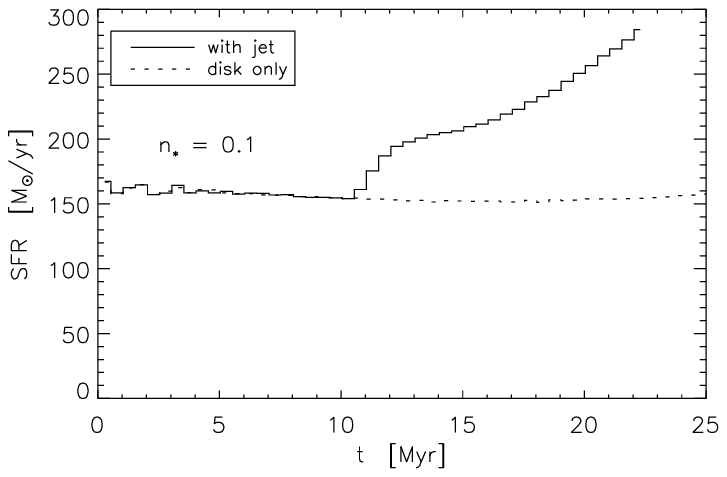}
   \includegraphics[width=0.9\linewidth]{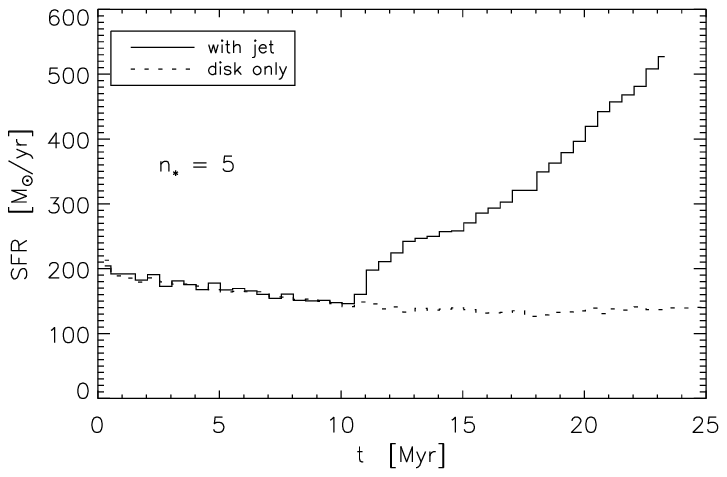}
   \caption{Total star formation rate for star formation thresholds of $n_\star = 0.1 \, \iccm$ (\emph{top}) and $n_\star = 5 \, \iccm$ (\emph{bottom}). The jet is started at $t = 10$ Myr.
   }
   \label{fig:total-sfr}
\end{figure}
\begin{figure}
   \centering
   \includegraphics[width=0.9\linewidth]{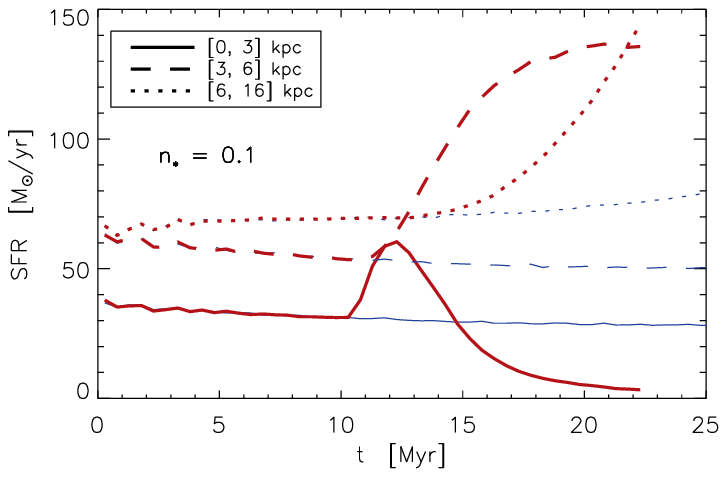}
   \includegraphics[width=0.9\linewidth]{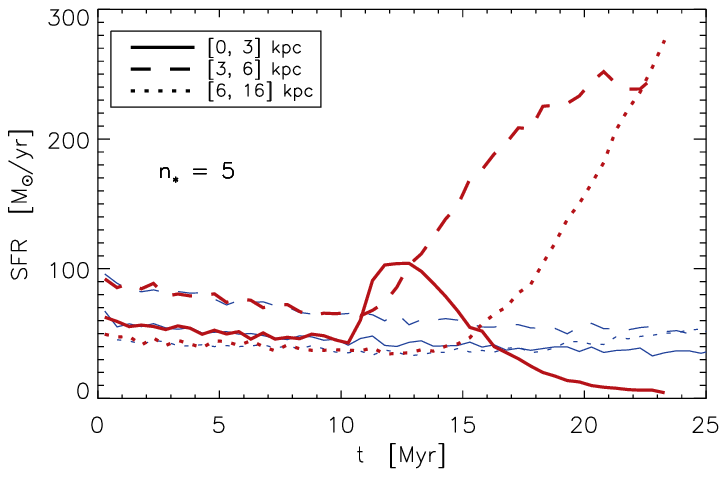}
   \caption{Evolution of the star formation rate for three cylindrical regions of the disc: the central area ($R < 3$ kpc), a ring-like region ($3$ kpc $\le R < 6$ kpc) and the outer disc ($6$ kpc $\le R < 16$ kpc). The \emph{top} panel is for a star formation threshold $n_\star = 0.1 \, \iccm$, the \emph{bottom} panel for $n_\star = 5 \, \iccm$. Only stars with $\Delta x < 10$ kpc from the disc plane are considered, the disc-only simulation is shown by \emph{blue, thin} lines, the \emph{red, thick} lines include the jet.
   }
   \label{fig:sfr-regions}
\end{figure}
\begin{figure*}
   \centering
   \includegraphics[width=0.41\linewidth]{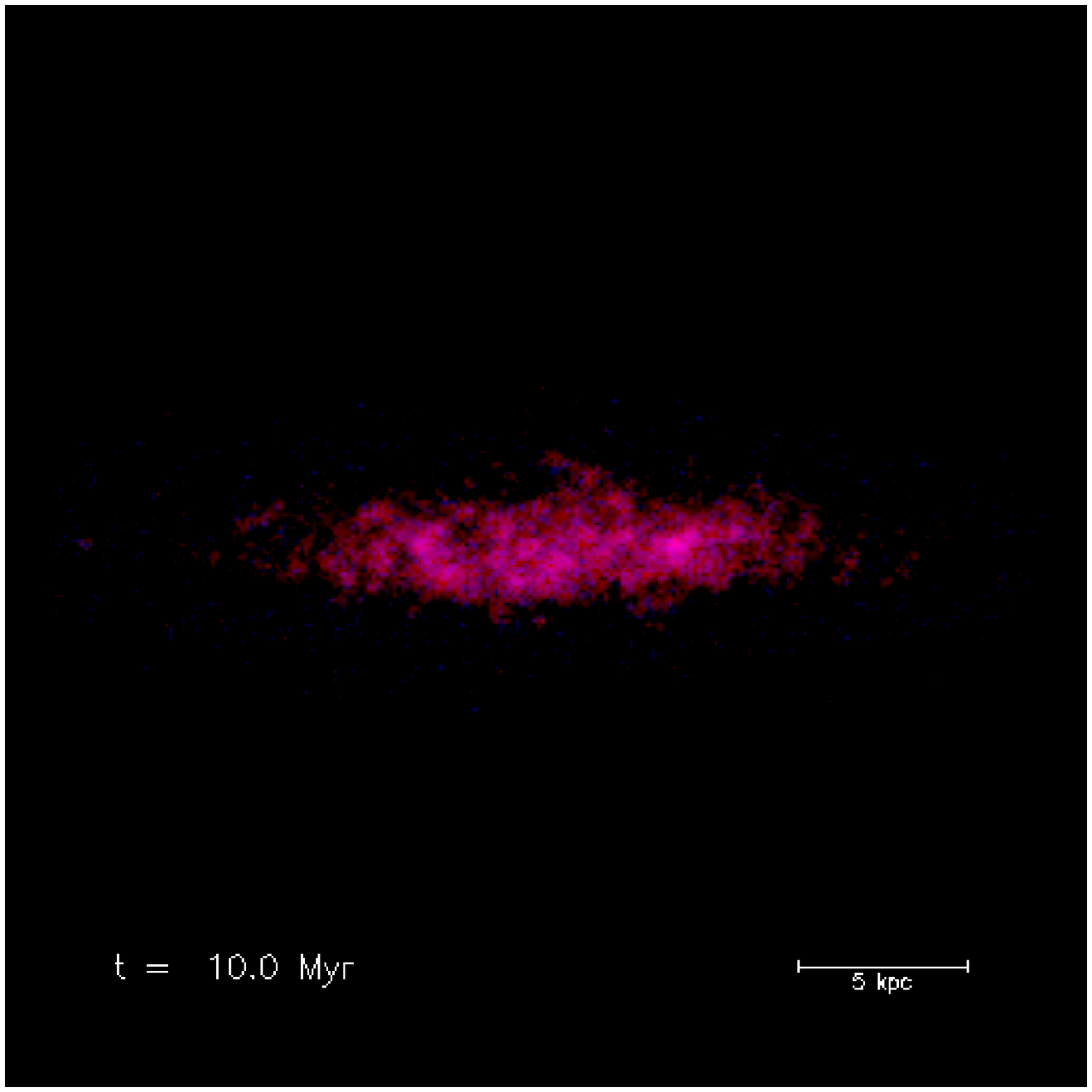}
   \includegraphics[width=0.41\linewidth]{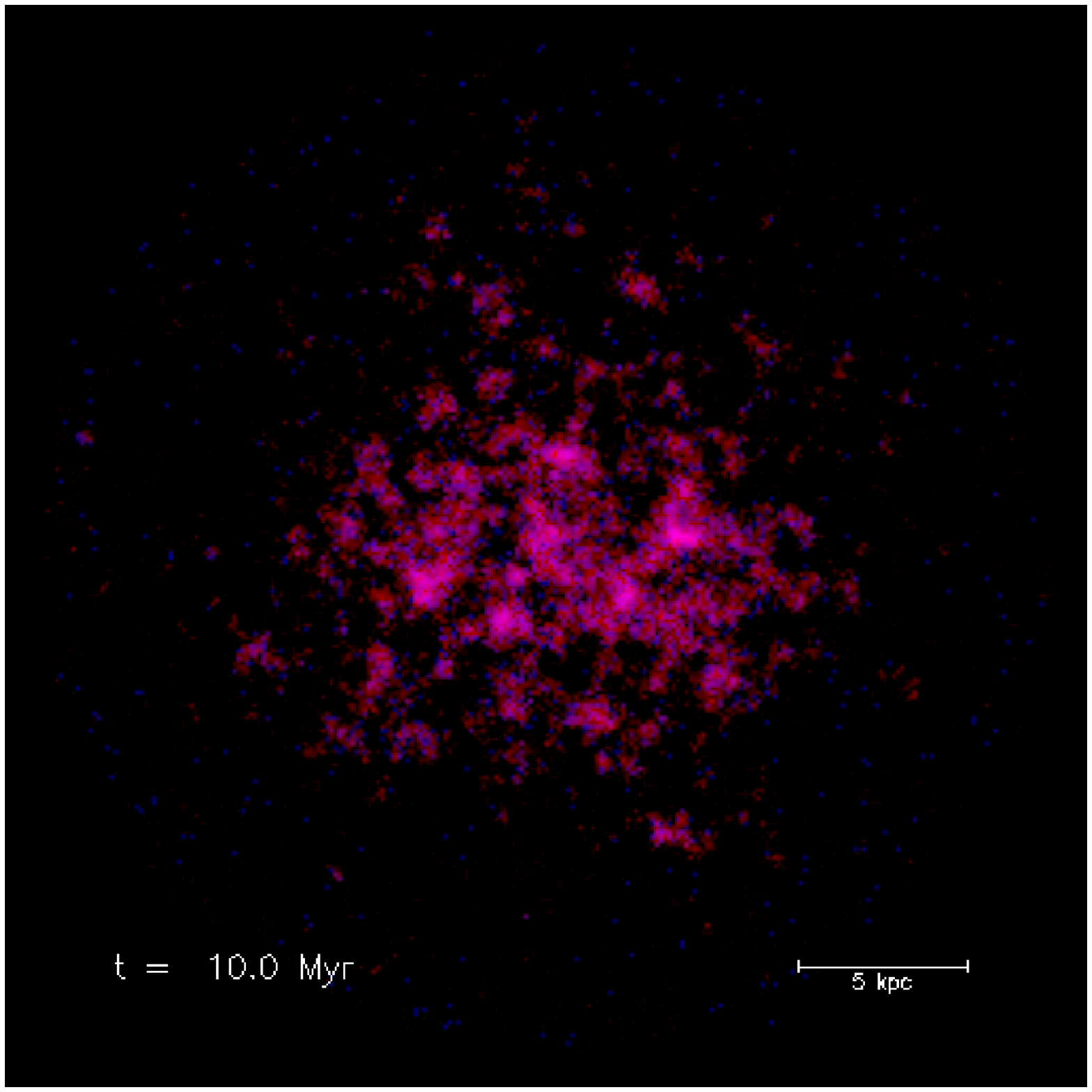}
   \includegraphics[width=0.41\linewidth]{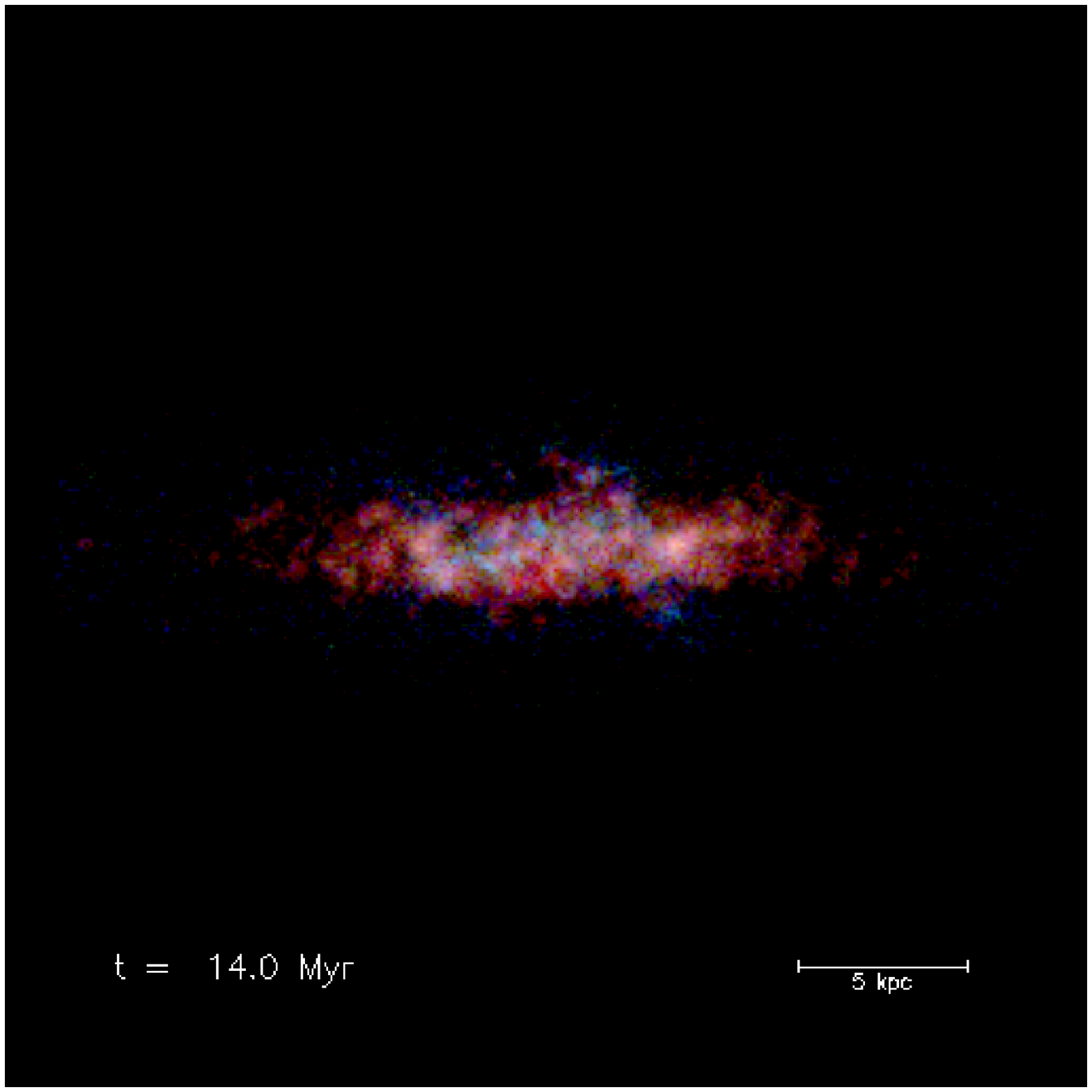}
   \includegraphics[width=0.41\linewidth]{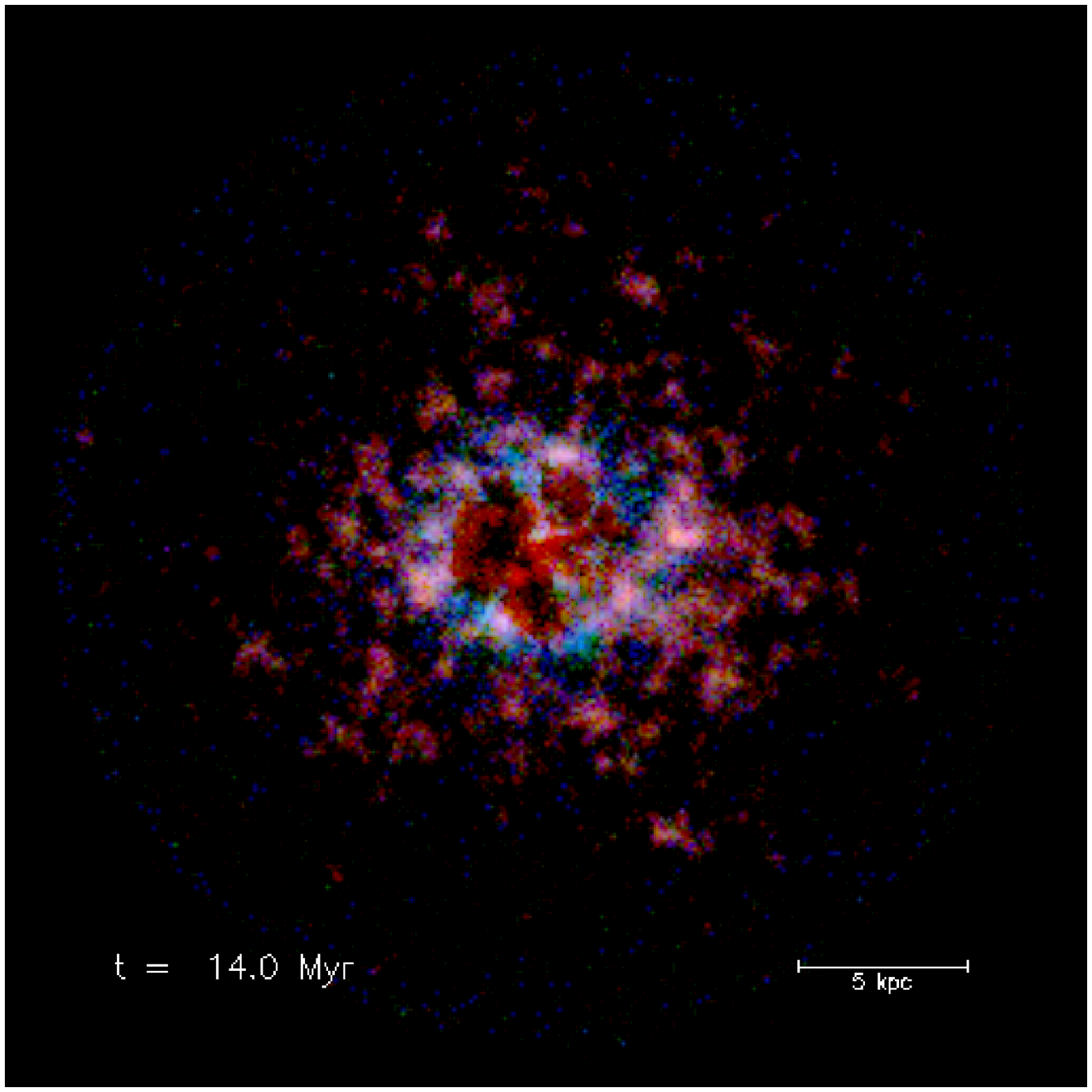}
   \includegraphics[width=0.41\linewidth]{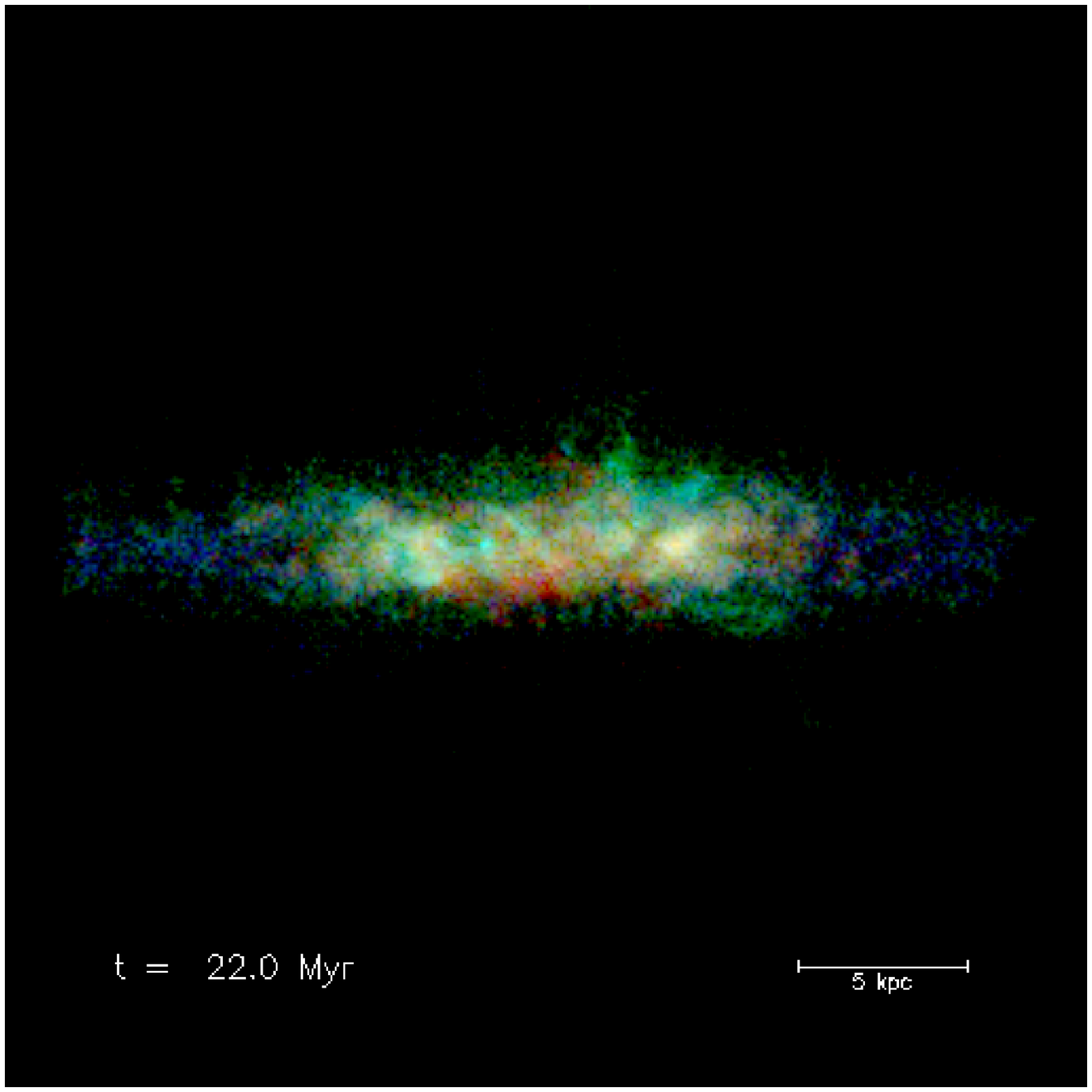}
   \includegraphics[width=0.41\linewidth]{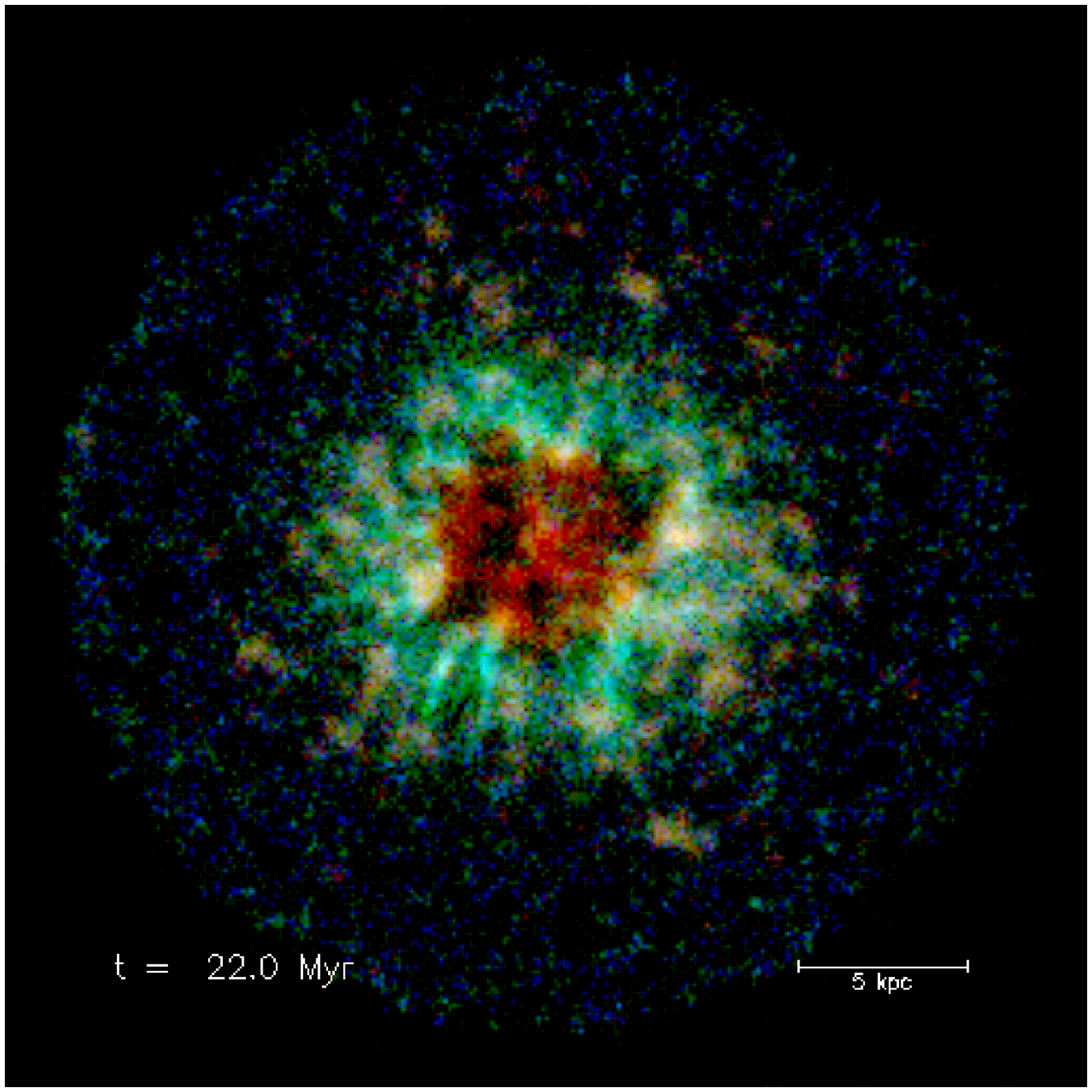}
   \caption{Star formation maps for $t = 10$ Myr (\emph{top}), $t = 14$ Myr (\emph{middle}) and $t = 22$ Myr (\emph{bottom row}) as RGB composite (\emph{left:} edge-on view, \emph{right:} face-on view). Newly formed stars (formed within the last 1 Myr) are shown in blue, all other are shown either in red (stars formed at $t \le 10$ Myr, in the undisturbed disc) or green (stars formed at $t > 10$ Myr, jet was active). The images refer to the simulation with the high density threshold ($n_\star = 5 \, \iccm$), a scale bar with $5$ kpc length is given.}
   \label{fig:sf-map}
\end{figure*}
The compression of gas by the blast wave and its short cooling times result in a considerable increase in the SFR despite the excavation in the centre (Fig.~\ref{fig:total-sfr}). The simulation with a higher star formation threshold ($n_\star = 5 \, \iccm$) shows a much stronger increase in star formation than the one with a lower threshold ($480 \, \Msunperyear$ vs. $280 \, \Msunperyear$ at $t = 22$ Myr). This can be understood as a consequence of the stronger cooling and compression at these potential sites of star formation. Also, the star formation is more concentrated towards the centre with the higher threshold due to the radial density profile. We have performed control simulations without the jets to ensure that the changes in star formation are actually caused by the jet activity, since the undisturbed disc cannot be in hydrostatic equilibrium due to the cooling. Indeed, the undisturbed disc evolves much more moderately than the disc with active jets. The relaxation effects in the disc are somewhat larger for the higher threshold run since stars are formed in smaller regions with higher densities, but they are generally opposite to the effects caused by the jet (they tend to lower the SFR). By the end of the simulation, the SFR is still increasing continuously. This can be understood by looking at the star formation in three different cylindrical regions around the disc axis (Fig.~\ref{fig:sfr-regions}). In addition, Fig.~\ref{fig:sf-map} shows face-on and edge-on maps of the stars at three different times.

\begin{figure*}
   \centering
   \includegraphics[width=0.45\linewidth]{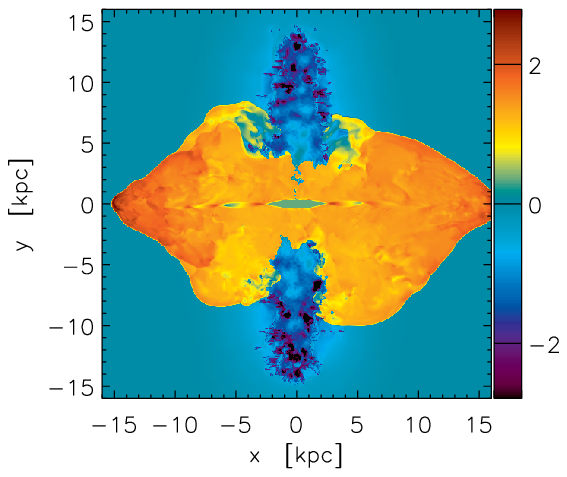}
   \includegraphics[width=0.45\linewidth]{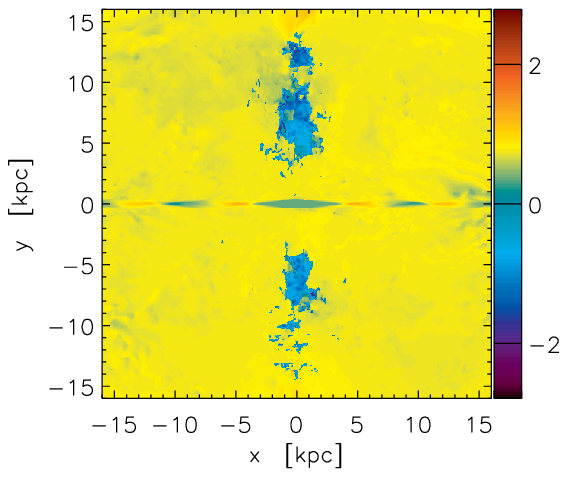}
   \caption{Pressure slices through $z = 0$ at $t = 13$ (\emph{left}) and $t = 22$ Myr (\emph{right}), showing $\log p$ normalized to the ambient pressure.}
   \label{fig:pressure-maps}
\end{figure*}
The SFR in the inner region ($R < 3$ kpc) drops after an initial increase caused by compression and the subsequent removal of gas. In the intermediate region, between 3 and 6 kpc, the SFR increases strongly by a factor of $2.5$ and $3.5$ for the low and high threshold, respectively, and then settles to this value since the main interaction region for the jet is then outside the disc and the increased SFR only continues eating up the generated high-density gas. The outer region ($R \ge 6$ kpc) only shows an increasing SFR at $t \ga 14$ Myr, but it continues to rise until the end of the simulation. The increase in SFR at large radii is generated by the overpressure in the jet's bow shock and cocoon, which is compressing the entire ISM of the disc (Fig.~\ref{fig:pressure-maps}) and thereby pushes its density beyond the star formation threshold. The control simulations show a slightly decreasing SFR except for the outer area, where it increases a bit since the ambient gas is already at slightly higher pressures than the cooling disc gas. So far, we cannot tell from the simulations how the long-term evolution of the SFR in the outer region is changed by the jet and how this depends on the modelling of the disc. Yet it is evident from the simulations, that this pressurizing of the disc occurs and may result in a large number of stars being formed.

For the overall impact of the jet activity, we identify three phases that explain the evolution of star formation in the disc: (a) excavation of a central cavity, (b) compression of disc gas and triggering of ring-like star formation by the blast wave and (c) late-time evolution of the disc in the jet cocoon. In the following subsections, we will address these phases in more detail. The phases should not be understood as consecutive phases, since they overlap temporally.

\subsection{Phase A: Formation of a central cavity}

The most direct consequence of the blast wave is the formation of a cavity in the disc centre since the bow shock pushes intermediate density gas outwards. The expansion can be approximated by a spherical shape, and assuming a constant energy input by the jet, the blast wave evolution can then be described by the one given in \citet{KrauseVLJ2}, keeping in mind the expectable deviations due to the inhomogeneities in the disc. The blast wave expands until it leaves the disc vertically, then venting its high pressure efficiently though this channel since the weaker density contrast there results in a faster jet propagation. Consequently, the lateral expansion of the cocoon into the disc slows down. As can be seen from the simulation, this breakout happens approximately when the bow shock reaches a radius of $r_\mathrm{cav} \sim 2 h_0$, where the vertical density profile dropped to $0.07$ of the midplane density. The time needed for this is, according to the blast wave approximation,
\begin{equation}
  t_\mathrm{cav} \sim 
  \left( 2 h_0 \right)^{5/3} \left( \frac{ 5 L_\mathrm{kin} }{ 4 \pi \rho } \right)^{-1/3} 
  \, ,
\end{equation} 
with $t_\mathrm{cav} \approx 3$ Myr for our simulation parameters and applying the mean central disc density as $\rho$. This means that for thinner discs and lower densities, the time-scales can be much shorter than 1 Myr, with $r_\text{cav} \ll 1$ kpc.

\begin{figure}
   \centering
   \includegraphics[angle=0,width=0.9\linewidth]{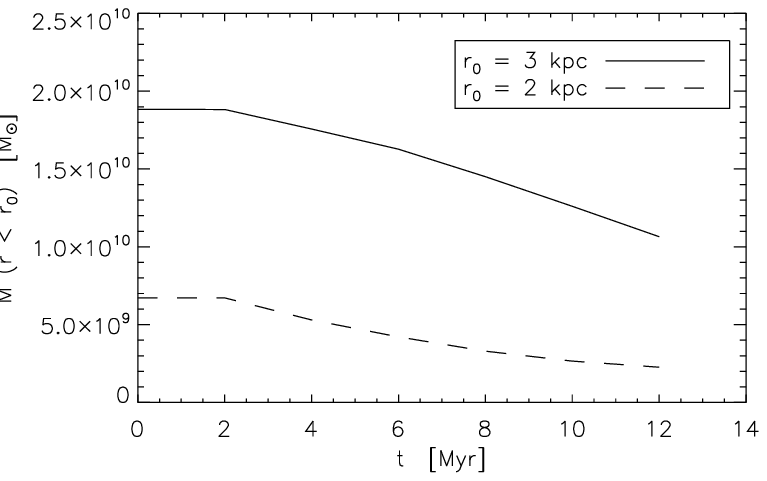}
   \caption{Evolution of the mass inside a spherical region $r < r_0$ around the
   centre with $r_0 = 2$ kpc and $3$ kpc. Initially overwriting the jet nozzle region accounts for $5.4 \times 10^8
   M_\odot$, and $< 10^9 M_\odot$ over the entire run time. }
   \label{fig:massinside-evol}
\end{figure}
The ``cavity'' actually is not empty but still contains a considerable amount of gas (Fig.~\ref{fig:massinside-evol}): about half the mass initially present in the centre is still there after the passage of the blast wave. However, the gas is located in very dense clumps and filaments with only a small filling factor.

We can furthermore estimate the energy deposited in this blast wave phase into removal of gas by assuming that half the mass of this volume is accelerated to a speed of $v = 0.5 \, r_\text{cav} / t_\mathrm{cav}$. 
With $M \sim 16 \pi h_0^3 \rho / 3$
, we get
\begin{equation}
  E_\text{kin} \sim 0.3 \, h_0^{5/3} \rho^{1/3} L_\mathrm{kin}^{2/3} \, ,
\end{equation}
which is in our case gives $\sim 5 \times 10^{58}$ erg. For comparison: the jet injects $5 \times 10^{59}$ erg within $t_\mathrm{cav} \approx 3$ Myr.

In our simulation, we start with an initial jets length of already $1.2$ kpc, so we actually have two blast waves originating from the tip of the jet beam, respectively. These blast waves meet near the disc midplane, get shocked by their respective counterparts and compress gas into a thin disc-like structure (easily visible in fig.~3 of \citetalias{Gaibler+2011}, which is immediately destroyed by the shear flows in the cocoon and fragments into the filaments and clumps that remain located in the cavity. If the imposed initial jet length were zero, the two blast waves would be merged from the beginning as a single blast wave originating from the disc centre, skipping the intermediate creation of a dense, double-shocked cavity disc. We can now estimate what can be expected in nature for this early blast wave formation.

The blast wave behaviour is a result of the jets having a much lower density than the ambient gas ($\eta \ll 1$), where then momentum balance enforces a slow propagation and most of the energy is deposited in thermal energy. In contrast, high density jets ($\eta \gg 1$) show a ballistic propagation behaviour. Hence, the blast wave phase will begin once the jet density drops strongly below the local ambient gas (i.e. once it encounters a region of much higher density). Larger jet beam widths are observed at larger distances from the galactic nucleus, stable jets showing only small opening angles of a few degrees. As jets expand slowly, their density decreases $\propto r^{-2}$ with distance (however, note that a Lorentz factor $\gg 1$ on the pc scale increases the effective density of the jet). Correspondingly, the jet might become underdense already at the scale of 10 pc to reach the strong density contrast we have at the kpc scale. These numbers are clearly very dependent on the jet properties and the gas distribution in the centre of the galaxy, but help putting this early stage into context. Once the two blast waves have merged, the evolution becomes independent of the initial blast wave locations. This consideration tells us that our setup is robust and despite the expected morphological differences at small scales, the results of this study would not be affected significantly.

Some disc gas gets expelled by the interaction with the jet, as can be determined by the compressible disc gas tracer field. A mass of the order of a few $10^8 \Msun$ on each side (for $|x| > 10$ kpc) is moved outwards from the initial disc ($|x| < 4$ kpc) with an asymmetric distribution (2 vs. $8 \times 10^8 \, \Msun$), as expected for the asymmetric interaction with the disc gas. By $t = 24$ Myr, it extends out to $\approx 20$ kpc above the disc plane. 
We note that the star formation at larger distances from the disc (more than 10 kpc above the disc plane) is ignored in our SFR -- it is generally low ($10^7 \, \Msun$ for $n_\star = 5 \, \iccm$ during the entire simulation).

\subsection{Phase B: Ring-like star formation}

While the blast wave creates a cavity in the disc centre, it pushes gas outward and compresses the disc at the cavity boundary with a ring-like or hour-glass shape. While the bow shock in vertical direction reaches out to lower densities, where the propagation becomes easier, it is much harder near the disc plane. The dense clumps in the disc with densities of $> 10 \, \mpccm$ have cooling times of $\lesssim 100$ yr at temperatures slightly above $T \sim T_\mathrm{min}$. Once they get shocked by the blast wave, they are compressed but lose their pressure support almost instantly, which prevents later re-expansion \citep[see also the simulations by][]{Mellema+2002,Fragile+2004}. This happens even after the bow shock has passed since the post-shock pressure is $~\sim 10^3$ times higher than initially in the disc, causing the clumps to collapse further.

The bow shock does not propagate very far into the disc due to the large mass in the disc and since the lateral expansion stalls once the jet has broken through the disc vertically. The strong compression of gas at the cavity boundary, however, leads to very efficient star formation with a ring-like morphology. Although our physical understanding of star formation is still limited, it is clear that high gas densities and a strong increase in environmental pressure have a strong impact on the actual star formation processes in molecular clouds \citep{Krumholz+2011}, which can not be modelled within this simulation.

\begin{figure}
   \centering
   \includegraphics[width=0.9\linewidth]{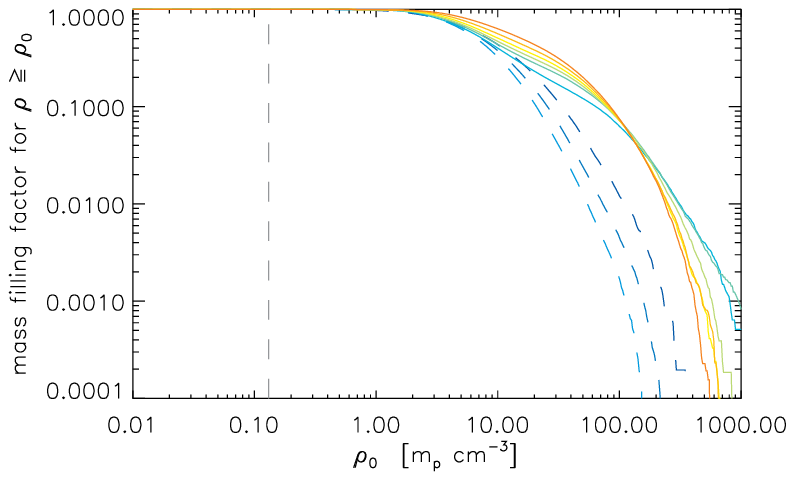}
   \includegraphics[width=0.9\linewidth]{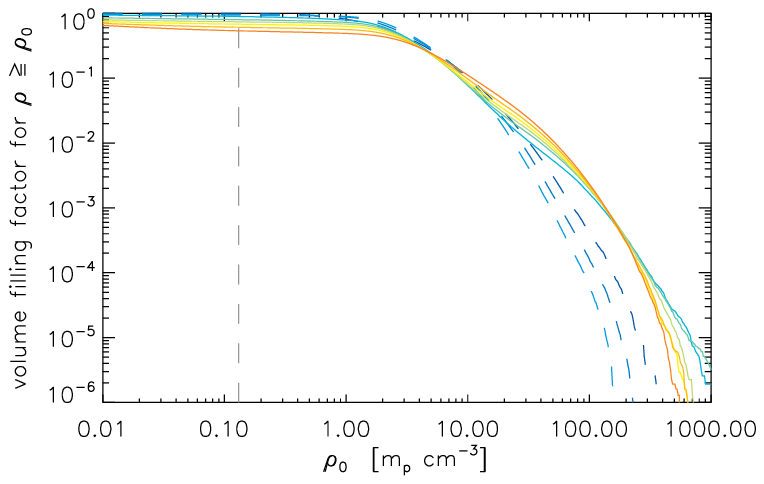}
   \caption{Evolution of the mass and volume filling factors of the disc. The time evolution is indicate by colour, going from blue via green and yellow to red. For early times (blue, \emph{dashed lines}, $t = 0$, 5 and 10 Myr), when the jet is still off, the filling factor at $\rho \gtrsim 10 \, \mpccm$) decreases. With an active jet (\emph{solid lines}, $t = 12$, 14, 16, 18, 20 and 22 Myr) it quickly increases, but then decreases again for the highest densities. Please note the jump between $t = 10$ Myr and $t = 12$ Myr. The star formation threshold $n_\star = 0.1 \, \iccm$ is indicated as grey vertical line (\emph{dashed}). The disc is defined as the geometrical region, where the disc density profile is larger than $\exp(-2) = 13.5$ per cent of its central value.}
   \label{fig:fillingfactors}
\end{figure}
The strong increase in density can also be seen clearly in the filling factor of the dense gas (Fig.~\ref{fig:fillingfactors}). Before the jet is launched, the filling factor for high-density gas decreases since this phase at $T = T_\mathrm{min}$ has a higher pressure than the lower density gas and expands (this is the relaxation process). However, once the jet becomes active the filling factor increases by approximately one order of magnitude. The largely increased mass in dense regions then results in much larger star formation. 

\begin{figure*}
   \centering
   \includegraphics[width=0.6\linewidth]{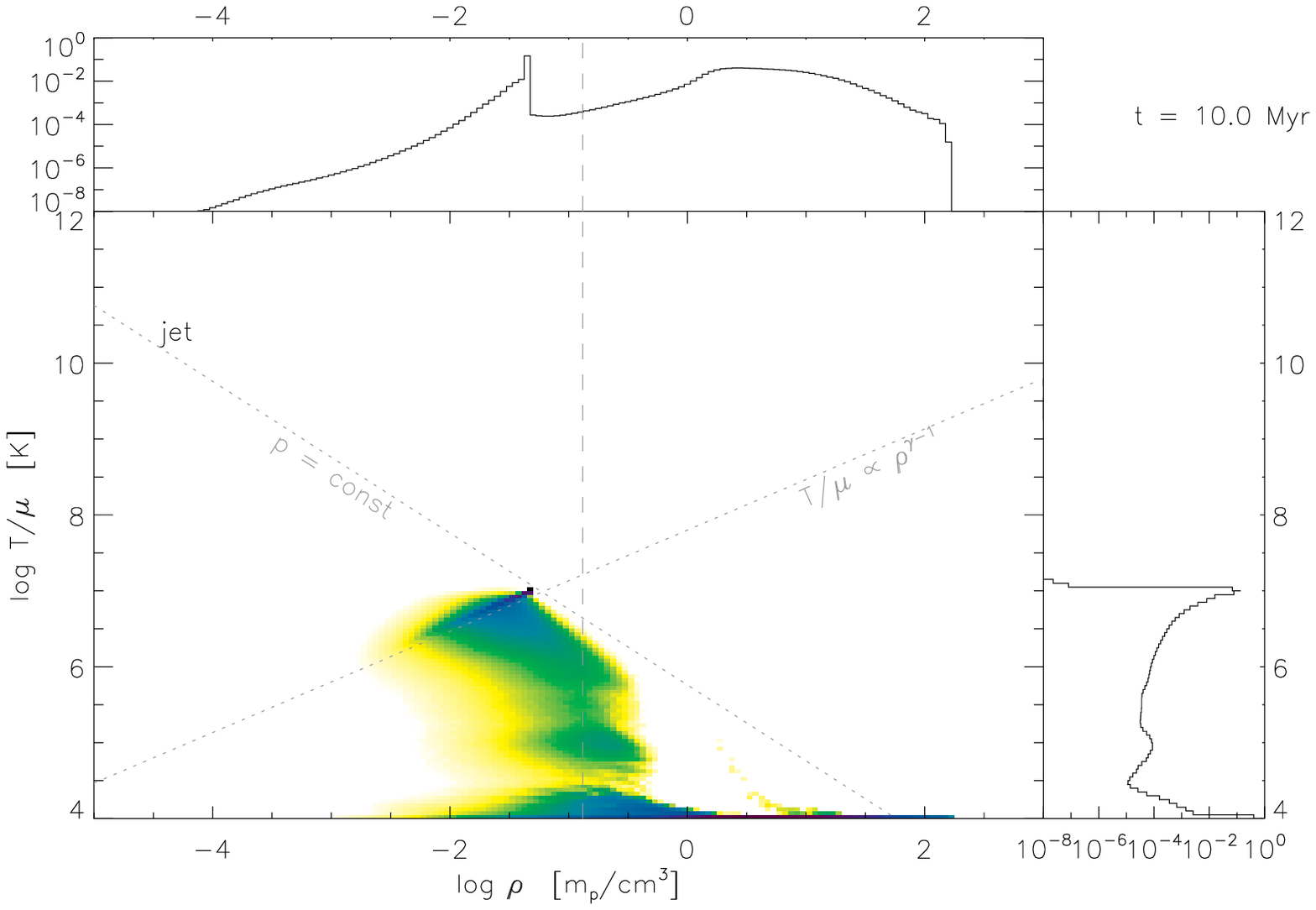}
   \includegraphics[width=0.6\linewidth]{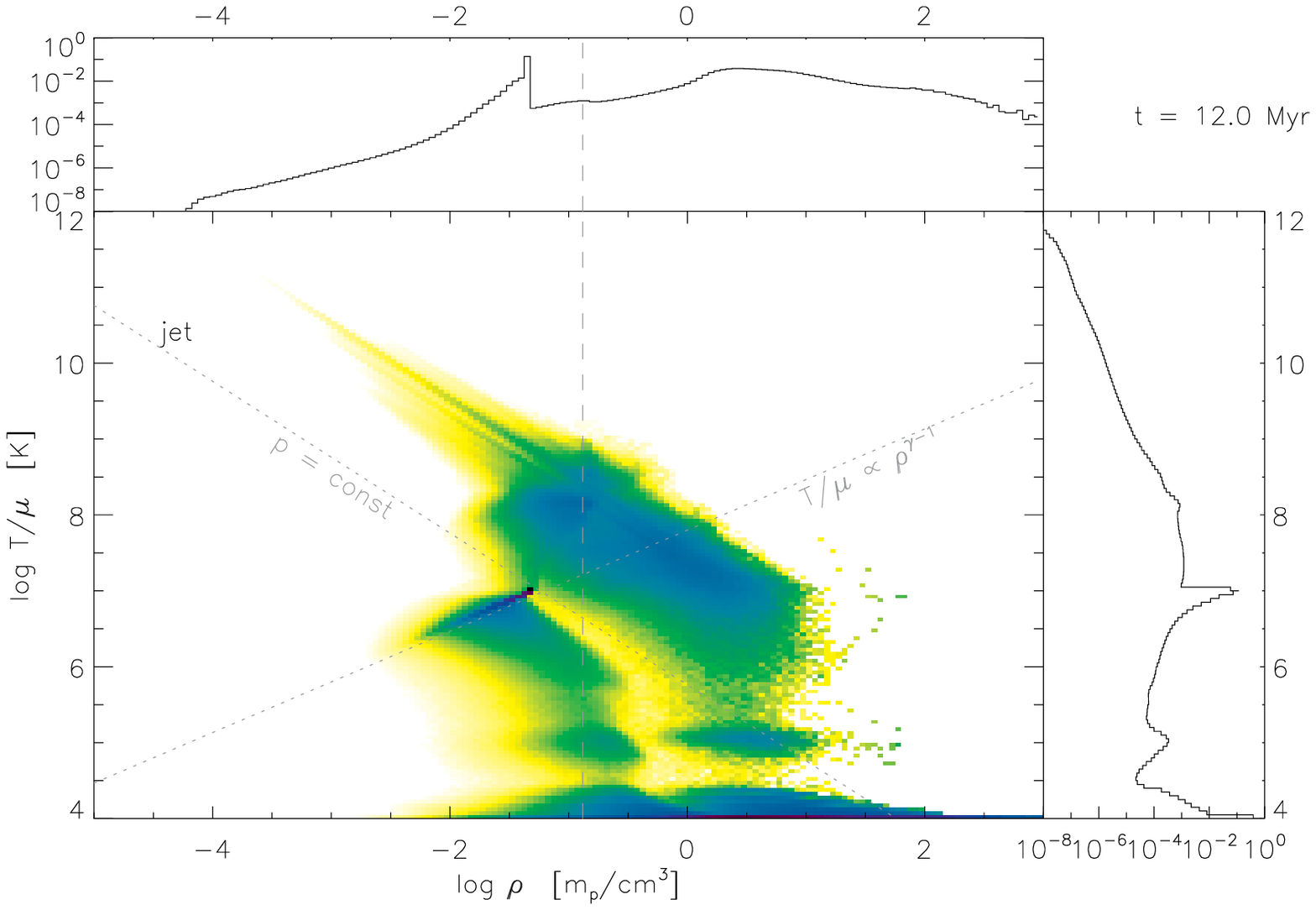}
   \includegraphics[width=0.6\linewidth]{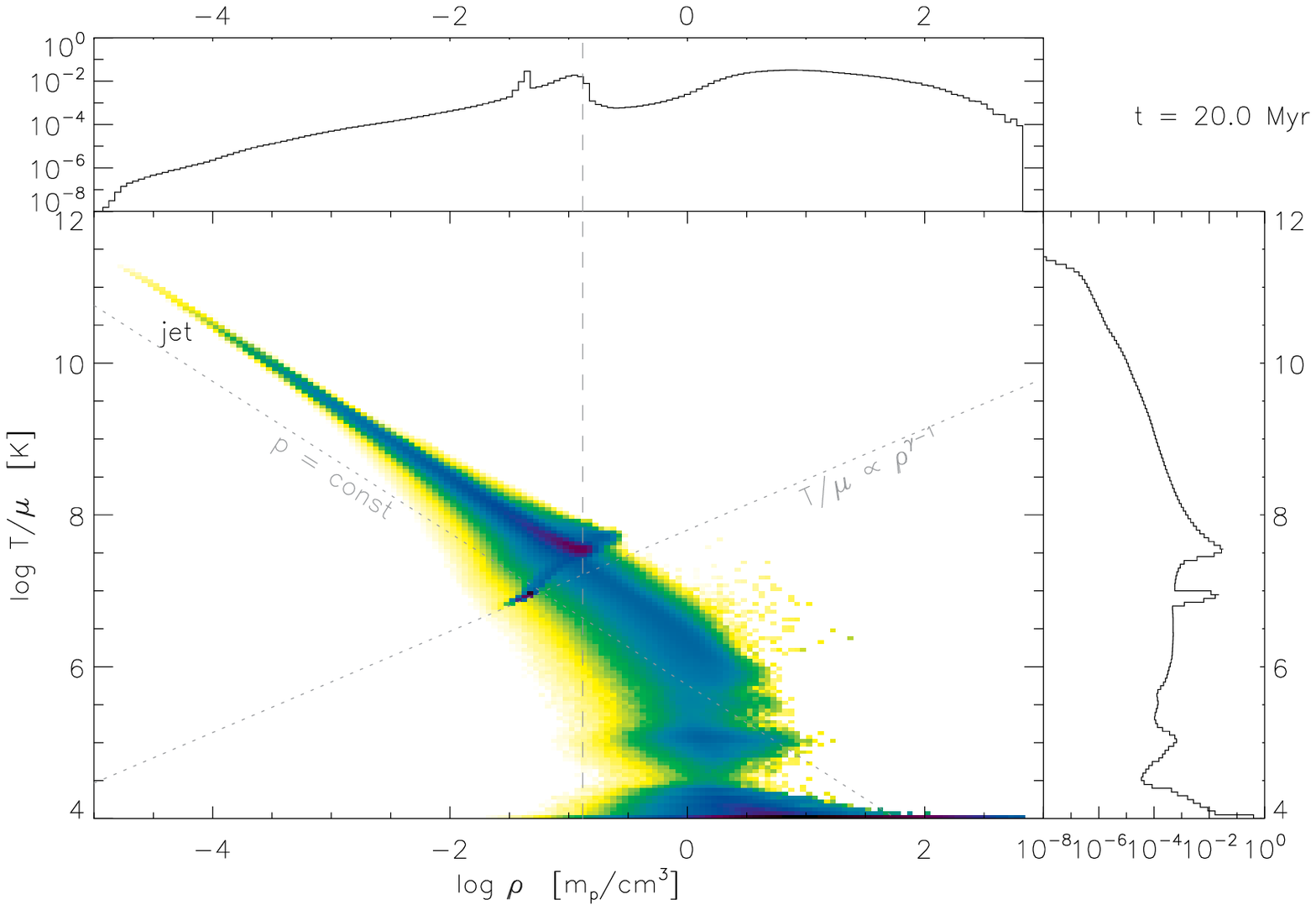}
   \caption{Phase diagram for the cells in the central $(32 \, \text{kpc})^3$ box, which 
   enclosed the entire disc: temperature $T/\mu$ vs. density $\rho$ for three different 
   times (the jet starts at $t = 10$ Myr. For comparison, lines of constant pressure as 
   well as adiabatic expansion/compression ($T \propto \rho^{\gamma-1}$) are overplotted (\emph{dotted}), the star formation threshold $n_\star = 0.1 \, \iccm$ is shown as vertical line (\emph{dashed}) and the location of the jet nozzle is indicated. Mass-weighted density and temperature histograms with logarithmic bins are attached to the phase diagram, respectively.}
   \label{fig:phasediagram-central}
\end{figure*}
The phase diagrams at three different times shown in Fig.~\ref{fig:phasediagram-central} give a different view of the interaction of the jet with the dense and dilute disc phases. Two lines, indicating the path for adiabatic expansion and for constant pressure, are overplotted. Gas cooling to $T = T_\mathrm{min}$ inevitably causes pressure gradients within the disc, resulting in the hot and the very dense phase of the ISM being at higher pressure than the intermediate density gas in the disc and consequently expanding during the early relaxation phase. Without additional heating processes (note that supernova feedback is not enough to balance the large radiative losses), no dynamical equilibrium can be established to keep the disc stable. We deliberately have chosen to neglect these processes since we cannot model them in a physically correct way and our setup without gravity exhibits only weak relaxation during the simulated time. At $t = 10$ Myr, only little gas is left at intermediate temperatures. Once the jet is active ($t = 12$ Myr), the jet cocoon appears as a new region at higher pressures. It extends from high temperatures, overpressured with respect to the jet nozzle and the ambient gas, down to lower temperatures and high densities. The latter part is the interaction and mixing zone between the dense disc gas and the very low density jet plasma. Here, the cooling times are short and the gas cools down quickly. There is a clear accumulation of gas at $T/\mu \approx 10^5$ K (where cooling times become somewhat larger) before it ends up quickly with $T = T_\mathrm{min}$ and underpressured, then reaching up to higher densities.
At $t = 20$ Myr, the cocoon is less overpressured and visible as a distinct $p = \mathrm{const}$ feature. Now the interaction with the ambient gas is more prominent and the gas distribution forms a downward-bent cooling path, which supplies the disc with an increased amount of dense, star-forming gas.

\subsection{Phase C: Later evolution in the jet cocoon}

At later times, when the jet mostly interacts with the ambient gas, the changed environment of the gas disc becomes important: the bow shock now encloses the entire disc and the increased pressure of the jet cocoon affects the disc. This is due both to the ram pressure of the backflow with its vortices as well as the thermal pressure of the cocoon. Since the cocoon turbulence is in the subsonic or transonic regime, the thermal pressure dominates somewhat over the ram pressure. The cocoon pressure now sets the environmental pressure for the gas disc, which gets compressed. In our simulation, this is mostly on the surface of the disc, but with time this pressure will affect the entire disc. There is also ablation at the disc surface by the cocoon plasma, however this is mostly intermediate density gas -- the high density, star-forming gas is much more robust and harder to affect, at sub-resolution scales it would also be stabilized by the self-gravity of the cloud.

Because the cocoon pressure drops as the radio source expands, the disc compression effect will eventually come to an end. \citet*{Gaibler+2009} find that jet cocoons are not strongly overpressured anymore once the source extends over more than $\sim 100$ kpc, where self-similar expansion then breaks down. We expect disc compression to be effective over approximately the active phase of the source or a bit longer. The total stellar mass formed by phase C is comparable to that of the previously described phase B.

\subsection{Stellar velocities}

\begin{figure}
   \centering
   \includegraphics[width=0.95\linewidth]{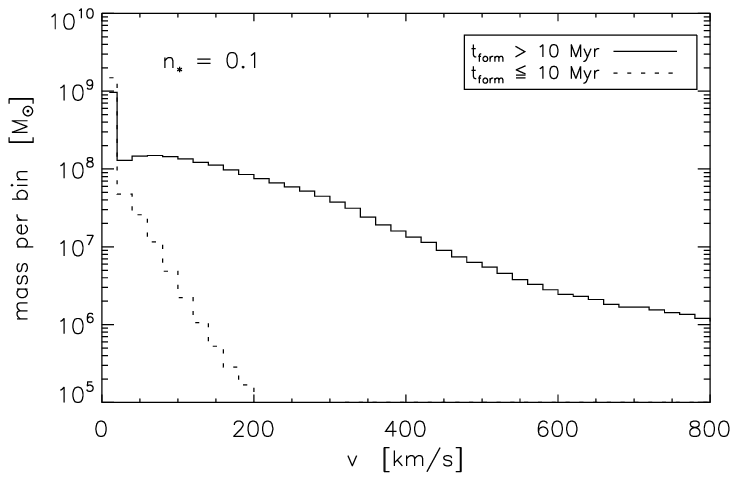}
   \includegraphics[width=0.95\linewidth]{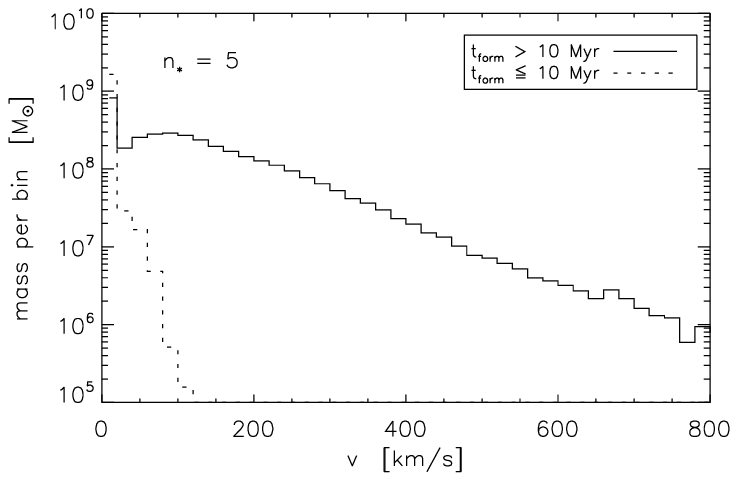}
   \caption{Velocity distribution of the stars at $t = 22$ Myr for the low ($n_\star = 0.1 \, \iccm$, \emph{top}) and high ($n_\star = 5 \, \iccm$, \emph{bottom}) star formation threshold. The contributions are shown separately for stars formed before (\emph{dotted}) and after (\emph{solid}) the jet starts at $t = 10$ Myr. The histograms show the mass of the stars within a bin size of 20 km s$^{-1}$.}
   \label{fig:v-stars}
\end{figure}
The dense gas is not only compressed by the blast wave and forms stars -- it is also accelerated by the ram pressure, mostly in the radial direction. This motion is imprinted on the newly formed stars within the clumps, which thus can show rather high radial velocities (Fig.~\ref{fig:v-stars}) of several $100$ km s$^{-1}$, even exceeding the escape velocity of the galaxy. The disc-only simulations, in contrast, show much lower velocities caused by the gas motions. This higher velocity dispersion may actually be an observable signature of these stars. While we expect this to occur also in nature, we caution that the present simulations are not sufficient for the determination of credible velocities because of two important points: first, the stars are not decelerated by gravitational forces since gravity is neglected in the setup, which is acceptable for the hydrodynamics on the short time-scales considered but has considerable impact on the stellar velocities. Second and even more important, stars form in the cores of molecular clouds at densities rather higher than the ones resolved here. Hence the real velocities of these cores are considerably lower than for the gas at densities corresponding to the star formation threshold $n_\star$. This is expected to depend on the internal structure of the clouds and the efficiency of momentum transfer. Accordingly, the measured velocities are upper limits. We note, however, that the velocity distributions for the two star formation thresholds considered are not drastically different. About 8.6 per cent ($n_\star = 0.1 \, \iccm$) or 8.3 per cent ($n_\star = 5 \, \iccm$) of the stars ($2 - 3 \times 10^8 \, \Msun$ in absolute numbers) formed at $t \ge 10$ Myr have velocities of $\ge 300$ km s$^{-1}$. Follow-up simulations with gravity and much larger resolution for the gas clouds will be needed to assess this aspect quantitatively.

\section{Discussion}
\label{sec:discussion}

\subsection{Consequences for galaxy evolution}

With the present simulations, we examine the impact of jets on a massive disc-like galaxy. The properties of the gaseous disc were chosen to be comparable with (non-merger) massive star-forming galaxies observed at redshifts of $z > 1$. The onset of the jets lead to an increase in SFR by a factor of two and more. During 12 Myr of jet activity, an additional stellar mass of $1.6 \times 10^9 \, \Msun$ ($7.2 \times 10^8 \, \Msun$) is formed for our runs with a star formation threshold of $n_\star = 5 \, \iccm$ ($0.1 \, \iccm$), respectively, compared to the quiescent disc simulation. Since the simulations are covering only $\sim 12$ Myr of jet activity, it is still difficult to estimate the long-term effects on the galaxy. As for the generally assumed negative feedback from jets, we have found that the reduction of star formation occurs mostly in the central region of the disc and is much smaller than the enhancement of star formation in the rest of the galaxy. By the end of the simulation, the SFR at intermediate radii does not increase further and would decrease slowly as the densest clumps are converted into stars. At larger radii, the SFR is still continuously increasing and is expected to stay at a high level until the jet cocoon pressure has dropped to a value near the original ambient pressure \citep[cf.][]{Gaibler+2009}, which happens after a few $10^7$ yr. Altogether, the total stellar mass generated additionally by the jet activity may be up to $\sim 10^{10} \, \Msun$. While this is only a fraction of the total stars formed considering the high quiescent (disc-only) SFR, these stars are distributed non-uniformly throughout the disc and are of similar age.

Both SFR-enhancing phases (B and C) are caused by the large amount of thermal energy generated by the jet as it forces its way though the ISM. Although direct interaction of the jet beam with clouds is  limited to a small volume, the resulting pressure affects a fair fraction of the disc at the early stage and eventually all of the galaxy once the bow shock has reached beyond the galaxy's radial extent. By that time (in our simulation within $\sim 10$ Myr), the galaxy is embedded in a hot and low-density atmosphere (the jet cocoon) with a pressure about one order or magnitude larger than initially. It is quite evident that increasing pressure in and around a galaxy within a relatively short time will have a considerable effect on its star formation, causing higher densities and higher star formation rates eventually -- independent of whether star formation rates are understood in terms of the local free-fall time (as done by the code), the Kennicutt--Schmidt law \citep{Kennicutt1998} or a global dependence on pressure \citep{WongBlitz2002,BlitzRosolowsky2004}.

{\changed
\citet{WagnerBicknell2011} note that the ablation time-scale in general would be shorter than the cloud collapse time-scale due to the large velocities in the jet cocoon, and destruction of the clouds hence would be expected without any possibly associated star formation. However, while ablation may actually remove a considerable fraction of the dense clouds, we expect that only the cloud envelopes with their lower densities would be affected. The star formation takes place only in the highest density regions where, at densities of $\sim 10^4 \, \iccm$ and considering stabilization by self-gravity, the jet will hardly have a totally destructive effect, in the worst case leaving a naked high-density core. Shocks driven into the clouds will be strongly radiative and dense shells formed by that would additionally protect the core even though the cloud may fragment \citep[cf.][]{Mellema+2002,Cooper+2009}. Since the environmental pressure of the cloud jumps to very large values, the cloud core will eventually be compressed to a higher density with an anticipated increase in star formation there. This might even happen in the cleared-out central disc region, where naked molecular cloud cores could be left embedded in the jet cocoon. This can be further quantified by the time scales of the processes competing with the cloud collapse. Following \citet{Cooper+2009} and using the data from the simulation and a fiducial radiative cloud of $100 \mpccm$ and $100$ pc radius, the cloud crushing time is $\gtrsim 10^8$ years and the growth time scale for Kelvin--Helmholtz instabilities is $10^5 - 10^6$ years. These are considerably larger than the cooling time for the dense clouds, and the Kelvin--Helmholtz time scale would even be larger if magnetic fields were present in the clouds. Hence the cloud would be stable against ablation until self-gravity takes over with a free-fall time scale of $\sim 3$ Myr and subsequent star formation would be expected.
}

If star-forming molecular clouds are seen as transient objects generated within a dynamic ISM of different phases, one could argue that destruction of warm (intermediate density) gas by the jet as in the cloud envelopes cuts off the supply necessary for the formation of molecular clouds and thereby would lower the star formation rate on longer time-scales. In the present simulations, however, the amount of strongly cooling gas is not changed much by the jet activity: depending on the selection threshold, it decreases only by 3 per cent for gas with cooling times $\tau_\mathrm{c} < 10^5$ yr or even increases by 10 per cent for $\tau_\mathrm{c} < 10^3$ yr between $t = 10$ and $t = 20$ Myr, hence providing a reservoir for molecular cloud formation also after the jet activity has ceased. 

{\changed
We have performed the simulation with two different star formation thresholds ($n_\star = 0.1 \, \iccm$ and $5 \, \iccm$) to examine resolution effects that are clearly expected for a setup with strong cooling. For both thresholds, the local Jeans length is resolved by $\geq 20$ cells, and it could even be raised to $100 \, \iccm$ for 4 cells per Jeans length. A stronger increase in SFR is found for the higher threshold, where star formation is more concentrated to the densest regions of the clouds, which we expect to be more realistic. Furthermore, the change in the density probability distribution function (PDF) caused by the onset of the jet activity is a robust and purely hydrodynamical result \citep[cf. compressional forcing,][]{Federrath+2010} with the direct consequence of a larger expected star formation rate independent of the actual implementation; the time evolution of the PDF is much larger than the differences between the runs with different thresholds at equal times. This increase in the density PDF at high densities is also found in the simulations by \citet[private communication]{WagnerBicknell2011}, with some differences due to the different setup. We are hence confident that the results are robust, and the enhancement of star formation effects might even be stronger if self-gravity of the gas clouds were considered.
}

Thermal or kinetic AGN feedback in cosmological and galaxy evolution simulations (modelling radio and/or quasar mode) is often thought to heat or expel gas from the galaxy and to be acting mostly on the circum-galactic gas 
\citep[e.g.][]{Springel+2005b,Croton+2006,Sijacki+2007,Bower+2008,BoothSchaye2009,Dubois+2011}, lowering the star formation rates. Our setup, in contrast, has shown the entirely opposite effect: the formation of an additional population of stars. The need for an additional boost of star formation has been recently suggested by \citet{KhochfarSilk2010}. They argue that to recover the high specific star formation rates that are observed in galaxies at $z>5$, $\Lambda$CDM models need to introduce stochastic boosts in star formation. Such boosts could indeed be triggered by jet activity as presented here in gas-rich galaxies at $z>5$. The main reason for the difference between our results and common AGN feedback simulations lies in the different treatment of the ISM and the collimated jets: whether it forms a rather continuous gas distribution or individual small-scale clouds are resolved and interact with the propagating jets. While on the AGN activity time-scale, our simulation shows a considerable positive feedback, this does not yet imply that there could be no negative feedback associated with jets: First, it is important to keep in mind that outside the galaxy, in the rather uniformly distributed and diffuse circum-galactic or intra-cluster gas, the feedback is still negative, heating the gas and quenching cooling flows. Second, on the galaxy scale, the formation of $\sim 10^{10} \, \Msun$ of stars (almost instantaneous compared to galaxy evolution time-scales and concentrated in certain regions, cf. Fig.~\ref{fig:sf-map}) will eventually be joined by the stellar feedback from these stars. This may have a strongly negative effect on the later evolution of star formation. 
{\changed For smaller gas masses, the positive effect might be very small and more efficient ablation might cause negative feedback to be dominant for the lower-density environment, potentially resulting in positive feedback in gas-rich (high-$z$) and negative feedback in gas-poor (low-$z$) environments. We note that while this study addresses the physics of ongoing jet feedback, it is still unclear under which circumstances AGN jets are launched and how frequently jet activity would occur in gas-rich galaxies.

Clearly the present simulations do not include all physics that might be relevant; magnetic fields and cosmic rays might play a role, and small-scale shocks or turbulent mixing layers may be effective below the finite resolution of our hydrodynamical simulation. However, these effects would affect discs without jets in a similar way unless the star formation mechanism is particularly sensitive to the changes. Assuming that giant molecular clouds are essentially self-regulated inside \citep[e.g.][]{Krumholz+2009}, it is at least not obvious why the molecular cloud processes in the core would change drastically, except that the hydrodynamical interaction with their environment changes the global cloud properties. In our simulations, compression by the blastwave and the jet cocoon in general is the dominant effect, with ablation and stirring of the cold clouds being limited to the surface of the disc. The long-term evolution of the galaxy and an exploration of the parameter space are beyond the scope of this paper and should be addressed in future studies.
}

\subsection{Comparison to observations}

{\changed 
As noted in Sect.~\ref{sec:introduction}, the observational evidence for positive feedback or jet-induced star formation is still limited, while the scenario in the light of the presented simulations appears almost inevitable from a theoretical point of view. Clearly, gas masses and jet powers in the local Universe vary over a large range and the present study only probes one point in a large parameter space (powerful jets in massive high-redshift gaseous discs). Accordingly, the star formation induced by the jets in the often quite different nearby objects will be on a considerably smaller level and possibly be unrecognized. We have seen in the simulation that the jet causes a morphological disturbance in the stellar distribution -- a ring or disc of stars is generated by the early blast wave of the jet expanding in the galaxy. Since the expansion becomes much slower and stalls once the jet is able to break out vertically through the dense gas, the radius of this stellar population should be on the order of a few vertical scale heights of the gas. For galaxies in the local Universe, this would be at some $100$ pc. Interestingly, \citet*{Jackson+1998} found a ring of young stars around the centre of Cygnus A, the most powerful radio source at low redshift ($z = 0.056$). 
They argue that the blue condensations are almost certainly due to young stars and not caused by scattering of the obscured quasar; the continuum energy distribution can be modelled by starburst activity much less than 1 Gyr ago, possibly an instantaneous burst with $t \leq 10 $ Myr, which is compatible with the radio source age of $\sim 10^7$ yr \citep{KrauseVLJ2}. \citet{Privon+2012} derive an overall SFR of $10$ to $70 \, \Msunperyear$ for Cygnus A. The morphology of the young stars is strikingly similar to the one found in our simulations. The radius of the stellar ring of $\sim 2$ kpc would mean that the ISM is rather extended in the vertical direction, but there are currently no observations available constraining the details of the distribution of a dense gaseous component.

Despite the small number of examples, the action of positive feedback is well compatible with or even supported by a number of observations mentioned in Sect.~\ref{sec:introduction}. In particular, the recent star formation activity in more than 75 per cent of small ($< 15$ kpc) radio galaxies \citep{Dicken+2012} or the young stellar populations found in radio galaxies \citep{Tadhunter+2002,Wills+2002,Holt+2007,BaldiCapetti2008,Tadhunter+2011} and the centres of radio galaxies \citep{Aretxaga+2001} may very well be caused by jet-induced feedback. And although the young stellar populations are mostly discussed in the context of merger-induced jet activity, it is not clear whether the merger scenario is a viable scenario and could explain the observations satisfactorily, whether it can be distinguished against jet-induced star formation and what effect jet activity would have while a merger is ongoing. The mere fact that the merger, increased star formation and radio jets occur at similar times does not give a causal relation, in particular since the duration of a single jet activity cycle is on the order of or even below the age accuracy of the young stellar population.
}

The additional stars in our simulation that have been formed due to the action of the jet are almost entirely located within the gaseous galactic disc. This may however not be interpreted as evidence against the possible jet induced star formation away from the disc, aligned with the axis of the radio jet \citep[e.g.][]{McCarthy+1987,Dey+1997}.
This possible mode of star formation is associated with tens of kpc sized emission line haloes \citep[e.g.][]{MeisenheimerHippelein1992,McCarthy1993,Best+1997b,MileyDeBreuck2008}, which are quite possibly due to gas extracted from the galaxy \citep{Nesvadba+2008,KrauseGaibler2010}.
In our simulation, we find that approximately $10^9 \, \Msun$ of disc gas leaves the galaxy during the time the jet is active, though this gas does not remain dense enough to form the observed emission line gas. The absolute number is rather encouraging, and should be regarded as a lower limit, since we neglect several important effects in this regard in our simulations: First, we do not simulate the initial phase of the jet until it reaches the extent of 1~kpc. This phase might result in substantial outward motions of the gas \citep{WagnerBicknell2011}. Second, the ISM is expected to be turbulent, which might also help to get it started from the disc. Third, the enhanced supernova feedback due to the jet-induced star formation within the disc should also add a considerable amount of kinetic energy to the gas. Once having left the disc, the gas would be further stirred up by interaction with the turbulent jet cocoon \citep{KrauseAlexander2007}. Also magnetic fields (neglected in our simulation) might help to stabilize the emission line phase. Finally, while generally speaking we simulate a high power \citep[FR II, ][]{FanaroffRiley1974}, the jet power in our simulation is set rather conservatively. If we would increase it by a factor of ten, and if the amount of ejected mass would scale linearly with the jet power, we would reach the right order of magnitude that has been claimed by the observations \citep{Nesvadba+2008}.
Since there is not even gas cold enough to form emission line regions present in our simulations outside the galactic disc, which is however clearly present in nature, it is evident, that our simulations do not address jet-induced star formation outside the galactic disc.

{\changed
Our simulations include a high-power FR II jet -- a case which is quite rare in the local Universe. The low-power (FR I) jets are much more abundant and evolve differently. Yet, the early stage may be very similar since FR I jets are still relativistic at small scales and would deposit their mechanical power into the ISM in a similar way as for the simulated case. Since their luminosities are much lower, their effect on the dense gaseous phase might be much smaller and restricted rather to small radii, which is possibly difficult to detect. In general, low-redshift radio galaxies show low star formation rates as derived from polycyclic aromatic hydrocarbon (PAH) emission features \citep{Shi+2007,Ogle+2010}, even though some of them have fairly large masses of molecular gas. Yet, we believe that a careful examination of radio galaxies for a central star formation and a global enhancement of star formation once the bow shock has propagated beyond the galaxy should be done, including formation of molecular gas at these locations, in order to shed more light on the impact of jet activity on the evolution of their host galaxies. Observations at high redshift are clearly more demanding. \citet{Stacey+2010} detected $158 \mu$m [CII] line emission in galaxies at $z \sim 1 - 2$, and inferred most of the FIR and [CII] emission originates from kpc-scale cores in the AGN-dominated galaxies where the local FUV field is much higher than in the star-formation-dominated galaxies. This led them to assert that the onset of AGN activity stimulates large-scale star formation, an argument also consistent with detection of strong 88 $\mu$m [OIII] and 122 $\mu$m [NII] emission in composite starburst AGN \citep{Ferkinhoff+2010,Ferkinhoff+2011}. \citet{Seymour+2008} were able to detect the PAH signature of a large star formation rate in a high-redshift radio galaxy. The wide range of observational results found for radio galaxies sources might indicate that positive feedback depends on certain conditions, and possibly be more effective for large disc masses which are more often found at higher redshift. 
}

\section*{Acknowledgments}
We thank the anonymous referee for constructive comments which helped to improve this paper.
VG wishes to acknowledge financial support by the Deutsche Forschungsgesellschaft (Priority Programme SPP 1177, ref. no. KH 254/1-1) and in part by the Sonderforschungsbereich SFB 881 ``The Milky Way System'' (subproject B4) of the German Research Foundation (DFG), support with RAMSES by Yohan Dubois and Romain Teyssier and an inspiring discussion with Bob Fosbury. 
SK and JS acknowledge support from the Royal Society Joint Projects Grant JP0869822. 
Computations were performed on the SFC cluster of TMoX at MPE and the Power6 VIP at RZG. 
MK acknowledges support by the DFG cluster of excellence ``Origin and Structure of the Universe''.
The authors made use of VAPOR for visualization purposes \citep{ClyneRast2005,Clyne+2007}.

\bibliographystyle{mn2e}
\bibliography{references}

\label{lastpage}

\end{document}